\documentclass[11pt,aps,prd,showpacs,nofootinbib,superscriptaddress,preprint,preprintnumbers,floats]{revtex4}

\pdfoutput=1

\usepackage[usenames,dvipsnames]{color}
\usepackage{graphicx}
\usepackage{mathrsfs}
\usepackage{amsmath,amsfonts,amssymb,dsfont,bm,dcolumn,float}
\usepackage{pstricks,feyn,verbatim}
\usepackage{url,hyperref}
\usepackage{ifthen}
\usepackage{bbm}

\usepackage{slashed,xcolor}
\usepackage{multirow,array,rotating}
\usepackage{fancybox,epstopdf}

\newcommand{\GeV}      {~\mathrm{GeV}}
\newcommand{\TeV}      {~\mathrm{TeV}}

\newcommand{\fb}      {~\mathrm{fb}}

\newcommand{\beqn}{\begin{eqnarray}}
\newcommand{\eeqn}{\end{eqnarray}}
\newcommand{\beqs}{\begin{subequations}}
\newcommand{\eeqs}{\end{subequations}}
\newcommand{\be}{\begin{equation}}
\newcommand{\ee}{\end{equation}}
\newcommand{\non}{\nonumber \\}

\newcommand{\mathsym}[1]{{}}

\def\gU{\rm U}
\def\gSU{\rm SU}

\def\mL{\mathcal{L}}

\def\mO{\mathcal{O}}

\def\hf{\frac{1}{2}}

\begin{document}

\title{LHC Searches for The Heavy Higgs Boson via Two B Jets plus Diphoton}

\author{Ning Chen}
\email{ustc0204.chenning@gmail.com}
\affiliation{Center for High Energy Physics, Tsinghua University, Beijing, 100084, China}
\author{Chun Du}
\email{chun.thazen.du@gmail.com}
\affiliation{Institute of High Energy Physics, Beijing, 100049, China}
\author{Yaquan Fang}
\email{fangyq@ihep.ac.cn}
\affiliation{Institute of High Energy Physics, Beijing, 100049, China}
\author{Lan-Chun L\"{u}}
\email{lvlc10@mails.tsinghua.edu.cn}
\affiliation{Center for High Energy Physics, Tsinghua University, Beijing, 100084, China}
\affiliation{Theoretical Particle Physics and Cosmology Group, Department of Physics,
King's College London, London WC2R 2LS, United Kingdom}

\begin{abstract}

Extra scalar fields are common in beyond Standard Model (SM) new physics, and they may mix with the 125 GeV SM-like Higgs boson discovered at the LHC. This fact suggests possible discovery channels for these new scalar fields with their decay modes involving the 125 GeV Higgs boson. In this work, we explore the LHC search potential of the heavy CP-even Higgs boson $H$ in the two-Higgs-doublet model. We focus on the channel of $H$ decaying to a pair of light CP-even Higgs bosons $h$, with two $h$'s decaying to two $b$ jets and diphoton sequentially. This channel is particularly involved when the relevant cubic coupling is enhanced. We find such enhancement to be possible when taking a large CP-odd Higgs mass input for the two-Higgs-doublet model spectrum. Analogous to the SM Higgs self-coupling measurement, the two $b$ jets plus diphoton final states are of particular interest due to the manageable SM background. After performing a cut-based analysis of both signal and background processes, we demonstrate the LHC search sensitivities for the heavy CP-even Higgs boson in a broad mass range via the two $b$ jets plus diphoton final states.

\end{abstract}


\maketitle


\section{Introduction}
\label{section:intro}

The $125\,\GeV$ Higgs boson discovered at the LHC $7\oplus 8\,\TeV$ runs~\cite{Aad:2012tfa, Chatrchyan:2012ufa} manifests the Higgs mechanism for the electroweak symmetry breaking (EWSB). The current data for different production channels and decay final states point to a very ``Standard Model (SM)-like'' Higgs boson with higher accuracies of the Higgs couplings to be obtained by the upcoming LHC $14\,\TeV$ run and the precision measurements at the ILC and TLEP projects~\cite{Peskin:2012we, Gomez-Ceballos:2013zzn}.

Besides the minimal one-doublet setup, it is generally possible to have the Higgs mechanism realized with the extended scalar sector for various motivations. Some typical examples include singlet scalar extensions~\cite{Morrissey:2012db, O'Connell:2006wi}, a second Higgs-doublet extension~\cite{Lee:1973iz, Lee:1974jb, Dimopoulos:1981zb, Chacko:2005vw, Mrazek:2011iu} (see Ref.~\cite{Branco:2011iw} for a recent review), and the Higgs-triplet extension~\cite{Konetschny:1977bn, Cheng:1980qt}, where the EWSB follows the usual $\gSU(2)_{L}\times \gU(1)_{Y}\to \gU(1)_{\rm em}$ pattern. For models with extended gauge symmetries such as $\gSU(2)\times \gSU(2)\times \gU(1)$, scalar fields such as Higgs bidoublet and/or Higgs triplets are often necessary for the spontaneous symmetry breaking, e.g., in the left-right symmetric model~\cite{Mohapatra:1977mj, Gunion:1989in, Perez:2008ha} and the three-site moose model~\cite{Abe:2012fb}. As a common feature among various extended scalar sectors, at least one extra neutral scalar field exists in their spectra, while other Higgs siblings carrying different quantum numbers indicate their real shapes. Thus the discovery of additional neutral scalar field would indicate the extension to the one-doublet Higgs scenario for the EWSB.

Our discussions here focus on the general two-Higgs-doublet model (2HDM), with the extra neutral scalar field referring to the heavy CP-even Higgs boson $H$. Comparing to the singlet field extension to the Higgs sector, the 2HDM, {\it per se}, is seemingly more complicated in its field contents. However, the singlet scalar fields can only mix with the SM Higgs doublet through the Higgs potential, which typically suppresses its coupling strengths with the SM fermions and gauge bosons. Hence, the searches for a singlet scalar at the LHC are generally challenging. As for the 2HDM, we consider the decay mode of $H\to hh$, with $h$ representing the light CP-even Higgs with mass of $125\,\GeV$. This decay mode arises from the cubic scalar coupling terms in the 2HDM potential, and it may become the most dominant one other than the conventional decay modes of $H\to (WW\,,ZZ\,,t\bar t)$ in certain mass range and parameter space. Distinct from the minimal supersymmetric SM (MSSM) Higgs sector where the $H\to hh$ decay mode can be dominant only in the mass window of $250\,\GeV\lesssim M_H\lesssim 2m_t$, we show the dominance of $H\to hh$ mode in a broad mass range of $H$, especially in the 2HDM-I case. Accordingly, the experimental searches for the $H\to hh$ mode may be considered also in addition to the conventional search strategies for $H$, namely, via the leptonic final states from the $H\to ZZ\to 4\ell$ and $H\to WW\to 2\ell 2\nu$ channels. Assuming the decay modes for the $125\,\GeV$ Higgs boson are very SM-like, one would expect the leading signal channels~\footnote{A general survey on the possible signal channels for a scalar field $S$ decaying into a pair of SM-like Higgs bosons is given in Ref.~\cite{Liu:2013woa}. } such as $ b \bar b b \bar b$, $b \bar b WW$ and $b \bar b \tau\tau$. Searches for these final states were studied for the SM Higgs self-coupling measurements~\cite{Dolan:2012rv, Papaefstathiou:2012qe, Goertz:2013kp, Barr:2013tda} and the beyond Standard Model (BSM) Higgs bosons~\cite{Dolan:2012ac, Kang:2013rj, No:2013wsa}. Additionally, the CMS Collaboration also carried out an analysis of the discovery potential of the extended Higgs sector with multilepton and photon final states at the LHC $8\,\TeV$ run recently~\cite{CMS:2013eua}. Although such final states are leading considerations because of the large branching ratios, they are typically challenging due to large QCD backgrounds, and a special technique of the Higgs jet substructure analysis~\cite{Butterworth:2008iy} is often required. Here, the rare decay channel of $ b \bar b \gamma\gamma$ is our primary interest for the heavy CP-even Higgs searches, in that the relevant SM background is under control. The cut-based analysis of both signal and background processes turns out to be sufficient for these final-state searches. Several previous studies on the SM Higgs boson self-coupling measurements also relied on the final states of $b\bar b\gamma\gamma$~\cite{Baur:2003gp, Baglio:2012np, Yao:2013ika, Barger:2013jfa}.

The paper is organized as follows. In Sec.~\ref{section:2HDM_Hhh}, we start with a review of the 2HDM, with emphasis on the $H\to hh\to b \bar b\gamma\gamma$ signal channel of our interest. Within the general 2HDM, we have the freedom of enhancing the partial width of $\Gamma[H\to hh]$. On the other hand, the conventional modes of $H\to ZZ\to 4\ell$ and $H\to WW\to 2\ell 2\nu$ are likely to be suppressed according to the current global fit to the 2HDM and become unfavorable for the experimental searches. In Sec.~\ref{section:2b2gamma}, we perform a cut-based kinematic analysis of the signal process of $pp\to HX$ with $H\to hh\to b \bar b\gamma\gamma$. The CP-even Higgs is studied in a broad mass range of $M_H\in (300\,,600)\,\GeV$, and the most optimal kinematic cuts for different $M_{H}$ inputs are obtained. Specifically, we probe the LHC search potential of the $(\alpha\,,\beta)$ parameter space for the $M_H=300\,\GeV$ case at the LHC run 2 and the high luminosity (HL) LHC runs. We also probe the mass reach for the $M_H\in (300\,,600)\,\GeV$ via the $b\bar b\gamma\gamma$ final states by restricting the 2HDM parameters in the regions that are consistent with the current global fit results. Finally we make conclusions and prospects in Sec.~\ref{section:conclusion}.


\section{The Exotic $H\to hh$ Channel in The 2HDM}
\label{section:2HDM_Hhh}

In the 2HDM, two complex Higgs doublets $\Phi_{1\,,2}$ carrying the hypercharge $Y=+1$ are introduced in the scalar sector. The most generic 2HDM potential is rich in its vacuum structures, while some simplifications are usually taken in practice. For our phenomenological studies below, we require a CP-conserving 2HDM potential. In addition, we assume the soft breaking of a discrete $\mathbb{Z}_{2}$ symmetry to keep the $m_{12}^2$ mass term, under which the Higgs doublets transform as $\Phi_1\to \Phi_1$ and $\Phi_2\to -\Phi_2$. The simplified 2HDM potential following these restrictions is expressed as
\beqn\label{eq:2HDM_potential}
V(\Phi_1\,,\Phi_2)&=&m_{11}^2|\Phi_1|^2+m_{22}^2|\Phi_2|^2-m_{12}^2(\Phi_1^\dag\Phi_2+h.c.)+\frac{\lambda_1}{2}(\Phi_1^\dag\Phi_1)^2+\frac{\lambda_2}{2}(\Phi_2^\dag \Phi_2)^2\non
&&+\lambda_3|\Phi_1|^2 |\Phi_2|^2+\lambda_4 |\Phi_1^\dag \Phi_2|^2 +\frac{\lambda_{5}}{2}\Big[(\Phi_1^\dag\Phi_2)^2+H.c.   \Big]\,,
\eeqn
where all parameters are real. Both Higgs doublets $\Phi_{1\,,2}$ acquire vacuum expectation values (VEVs)
\beqn\label{eq:2HDM_vevs}
&&\langle \Phi_{1} \rangle =\frac{1}{\sqrt{2}}\left( \begin{array}{c} 0 \\ v_{1} \\ \end{array}  \right)\qquad  \langle \Phi_{2} \rangle =\frac{1}{\sqrt{2}}\left( \begin{array}{c} 0 \\ v_{2} \\ \end{array}  \right)\,,
\eeqn
to trigger the EWSB, with the ratio of two Higgs VEVs to be parametrized as $t_{\beta}\equiv v_{2}/v_{1}$. After the EWSB, one is left with five scalars $(h\,,H\,,A\,,H^{\pm})$ in the physical spectrum by diagonalizing the mass terms in the potential (\ref{eq:2HDM_potential}). Two CP-even mass eigenstates $(h\,,H)$ are the mixtures of the scalar gauge eigenstates $(\rho_{1}\,,\rho_{2})$:
\beqn\label{eq:evenHiggs}
&&\left(\begin{array}{c} H \\ h   \end{array} \right)=\left(\begin{array}{cc} c_{\alpha} & s_{\alpha} \\ -s_{\alpha} & c_{\alpha}    \end{array} \right)\left(\begin{array}{c} \rho_{1} \\ \rho_{2}   \end{array} \right)\,.
\eeqn
In the following, the light CP-even Higgs boson $h$ in the 2HDM will be always considered as the one discovered at the LHC with mass of $125\,\GeV$~\footnote{Throughout this work, we will always consider the light CP-even Higgs boson $h$ to be the unique scalar with mass of $125\,\GeV$ in the 2HDM spectrum. The recent proposal of multiple Higgs bosons with degenerate mass~\cite{Gunion:2012he, Drozd:2012vf} is beyond the scope of our analysis here.}.

To avoid the serious tree-level flavor changing neutral current problem, one often enforces discrete symmetries to arrange the Yukawa couplings between each of the Higgs doublet $\Phi_{i}$ and the specific right-handed SM fermions. Our discussions will focus on two setups known as 2HDM-I and 2HDM-II, as listed in Table.~\ref{tab:2HDM_fermion}. Using the current data from the LHC $7\oplus 8\,\TeV$ runs, one could already constrain the 2HDM parameter space $(\alpha\,,\beta)$ by the global fit to the $125\,\GeV$ Higgs boson signal strengths. The sensitive final states involved for the global fit include the bosonic ones $h\to (\gamma\gamma\,,ZZ^{*}\to 4\ell\,,WW^{*}\to 2\ell 2\nu)$ together with the fermionic ones $h\to (b \bar b\,,\tau^{+}\tau^{-})$. Some facts are obvious for the light CP-even Higgs signal predictions in the 2HDM: (i) $h$ only decays into the SM final states lighter than itself, (ii) the $h$ coupling terms with SM fermions and gauge bosons are solely controlled by the parameters $(\alpha\,,\beta)$, and (iii) the only new particles contributing to the $h$ decay modes in the 2HDM are the charged Higgs bosons $H^{\pm}$ through the $h\to \gamma\gamma$ triangle loop, whose effects are regarded negligible when taking the large $M_{H^{\pm}}$ limits~\footnote{A large mass input of $M_{H^{\pm}}$ is also required by taking the ${\rm Br}[b\to s\gamma]$ as the constraint, with $H^{\pm}$ regarded as the only particles inducing this rare decay mode at the loop level~\cite{Neubert:2004dd}.}. The details of fitting the $125\,\GeV$ Higgs boson signal strengths within the 2HDM can be found in Refs.~\cite{Djouadi:2013vqa, Coleppa:2013dya, Chen:2013rba,Craig:2013hca, Barger:2013ofa}. Consistent with the current experimental data, the global fit pointed to the so-called ``alignment limit''~\footnote{The limit of $c_{\beta-\alpha}\to 0$ is also termed as the ``decoupling limit'' in literatures. Here we use ``alignment limit'' to highlight that the light CP-even Higgs boson is SM-like with the global fit to its signal strengths.} where $c_{\beta-\alpha}\to 0$. Consequently, one has $g_{hVV}\to g_{hVV}^{\rm (SM)}$ and $g_{h f f}\to g_{h f f}^{\rm (SM)}$ under this limit. In our analysis, we often take the following alignment parameter sets:
\beqn\label{eq:align_para}
&&{\rm 2HDM-I}:c_{\beta-\alpha}=0.4\,,~~{\rm 2HDM-II}:c_{\beta-\alpha}=-0.02\,,
\eeqn
and vary $t_{\beta}\in(1\,,10)$. The parameters of Eq.~(\ref{eq:align_para}) are chosen to be consistent with the $95\,\%$ C.L. of the $125\,\GeV$ SM-like Higgs boson signal fitting in the 2HDM parameter space $(\alpha\,,\beta)$~\cite{Craig:2013hca}. We also fix two other input parameters of $M_{A}=600\,\GeV$ and $\lambda_{5}=-6$ subject to the Higgs potential stability constraints, which was checked by the {\tt two-Higgs-doublet model calculator}~\cite{Eriksson:2009ws}. The perturbative unitarity and the electroweak precision constraints are not taken into account in our evaluations. These constraints receive contributions not only from the Higgs sector of the 2HDM but also from the BSM new physics contributions. For this reason, such constraints are less necessary for our following phenomenological discussions on the heavy CP-even Higgs.

\begin{table}[t]
\centering
\begin{tabular}{c|c|c|c}
\hline
 Models &  $u_{i\,,R}$ & $d_{i\,,R}$ & $\ell_{i\,,R}$   \\\hline
 2HDM-I & $\Phi_{2}$ & $\Phi_{2}$ & $\Phi_{2}$   \\
 2HDM-II & $\Phi_{2}$ & $\Phi_{1}$ & $\Phi_{1}$  \\
\hline
 \end{tabular}
\caption{The Yukawa coupling setups for the 2HDM-I and 2HDM-II fermion contents. The subscript $i$ denotes the generational indices.}\label{tab:2HDM_fermion}
\end{table}

\begin{table}[t]
\centering
\begin{tabular}{c|c|c|c}
\hline
 Models &  $\xi_{H}^{u}$ & $\xi_{H}^{d}$ & $\xi_{H}^{\ell}$   \\\hline
 2HDM-I & $s_{\alpha}/s_{\beta}$ & $s_{\alpha}/s_{\beta}$ & $s_{\alpha}/s_{\beta}$   \\
 2HDM-II & $s_{\alpha}/s_{\beta}$ & $c_{\alpha}/c_{\beta}$ & $c_{\alpha}/c_{\beta}$  \\ 
\hline
 \end{tabular}
\caption{The Yukawa couplings (normalized to the SM ones) of the SM fermions to $H$ in 2HDM-I and 2HDM-II.}\label{tab:2HDM_Hyukawa}
\end{table}

Next we list the relevant coupling terms for $H$. The tree-level couplings with the electroweak gauge bosons $V=(W\,,Z)$ are rescaled from the SM couplings by
\beqn\label{eq:H_gauge}
&&g_{HVV}=c_{\beta-\alpha} g_{HVV}^{\rm (SM)}\,,\qquad g_{HVV}^{\rm (SM)}=\frac{2 M_{V}^{2}}{v} \,.
\eeqn
The Yukawa couplings for $H$ depend on the model setups, which are generally expressed as
\beqn\label{eq:H_Yukawa}
-\mL_{Y}^{H}&=&\sum_{f}\frac{m_{f}}{v}\xi_{H}^{f} H \bar f f\,,
\eeqn
with the dimensionless factors $\xi_{H}^{f}$ presented in Table.~\ref{tab:2HDM_Hyukawa} for both 2HDM-I and 2HDM-II. From the general 2HDM potential (\ref{eq:2HDM_potential}), we obtain a cubic $Hhh$ coupling term:
\beqn\label{eq:Hhh_coup1}
\lambda_{Hhh}&=&\frac{c_{\beta-\alpha}}{v}\Big[(3M_{A}^{2}+3\lambda_{5} v^{2} -2M_{h}^{2}-M_{H}^{2})\non
&&\times \Big( c_{2(\beta-\alpha)}-\frac{ s_{2(\beta-\alpha)} }{t_{2\beta} } \Big)-M_{A}^{2}-\lambda_{5}v^{2}  \Big]\,,
\eeqn
where we trade the quartic Higgs self-couplings $\lambda_{1\,,...,4}$ into the 2HDM mass parameters for convenience. Around the alignment limit, this coupling of Eq.~(\ref{eq:Hhh_coup1}) can be approximately expressed as:
\beqn\label{eq:Hhh_approx}
\lambda_{Hhh}&\approx&-\frac{c_{\beta-\alpha}}{v}\Big[ 4M_{A}^{2}+4\lambda_{5}v^{2} -2M_{h}^{2} - M_{H}^{2}+\mO(c_{\beta-\alpha})   \Big]\,.
\eeqn
One feature of the general 2HDM is that $(M_{A}\,,\lambda_{5})$ are essentially free parameters. In contrast, this cubic coupling $\lambda_{Hhh}$ in the MSSM is reduced to
\beqn\label{eq:Hhh_MSSM}
&&\lambda_{Hhh}^{\rm MSSM}=-\frac{m_{Z}^{2}}{v}\Big( 2s_{2\alpha} s_{\alpha+\beta}-c_{2\alpha}c_{\alpha+\beta}  \Big)\,,
\eeqn
if one takes the tree-level MSSM Higgs mass relations by neglecting the radiative correction term from the squark loops. By this fact, the decay mode of $H\to hh$ for the MSSM case is at most attractive in the mass window of $250\,\GeV\lesssim M_{H}\lesssim 2m_{t}$. For the larger $M_{H}\gtrsim 2m_{t}$ regions, the decay mode of $H\to t \bar t$ is guaranteed to be the most dominant one due to the large top-quark Yukawa coupling. On the other hand, it is likely to have this cubic scalar coupling $\lambda_{Hhh}$ magnified to a significant amount once one abandons the MSSM Higgs mass relations, as is the case in the general 2HDM.

\subsection{The decay modes of $H$}

\begin{figure}
\centering
\includegraphics[width=6.8cm,height=4.5cm]{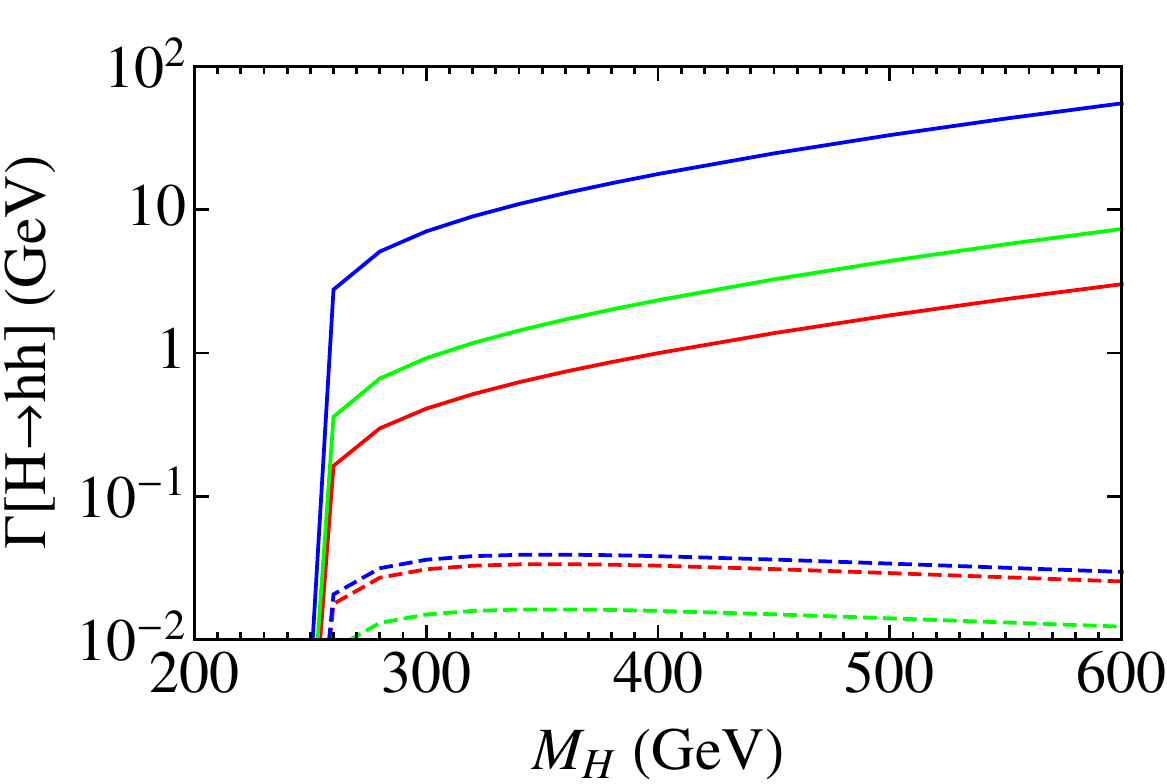}
\includegraphics[width=6.8cm,height=4.5cm]{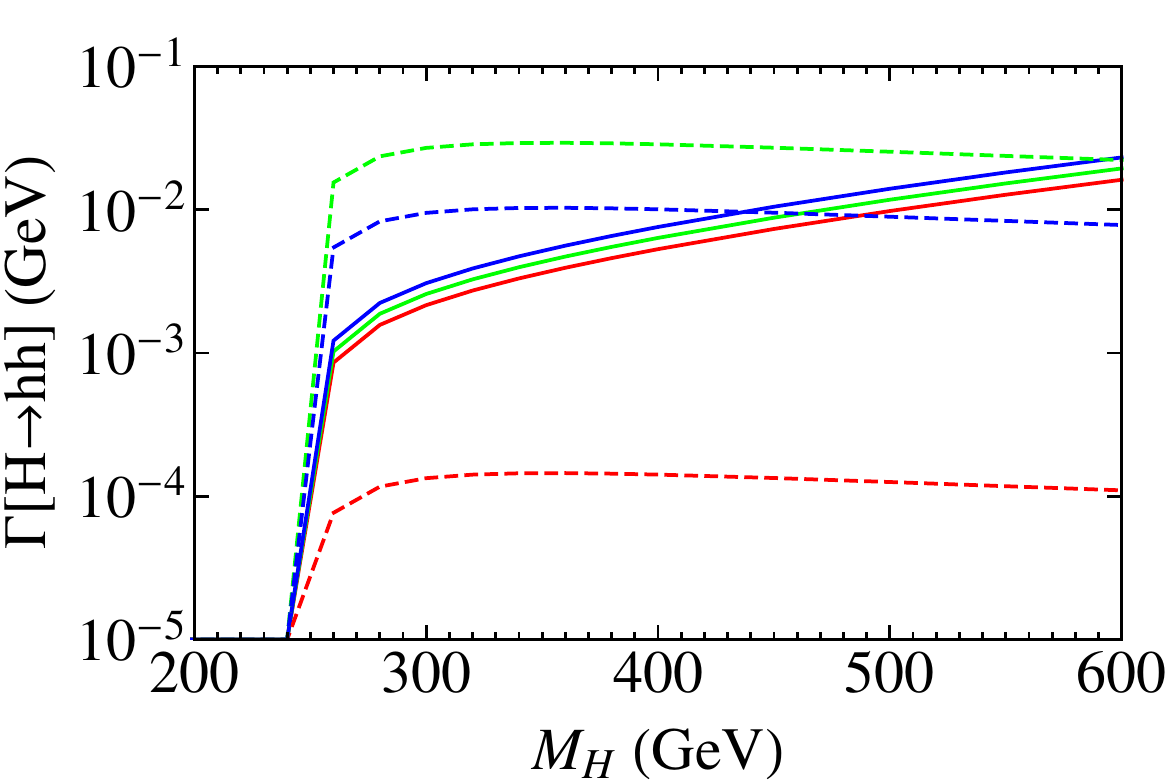}
\caption{\label{fig:Hhh_wid} The partial decay width of $\Gamma[H\to hh]$ for the $M_{H}\in (200\,,600)\,\GeV$ mass range, (left) 2HDM-I and (right) 2HDM-II. The partial decay widths $\Gamma[H\to hh]$ are demonstrated for both the MSSM-like case (dashed curves) and the general 2HDM case (solid curves) in both plots, together with different inputs of $t_{\beta}=1$ (red), $t_{\beta}=5$ (green), and $t_{\beta}=10$ (blue). }
\end{figure}

A heavy CP-even Higgs decays into the SM final states of $H\to (f\bar f\,, WW/ZZ)$ at the tree level, with the relevant partial decay widths obtained by rescaling from the corresponding SM Higgs cases:
\beqs\label{eq:Hdecay_SMtree}
\beqn
\Gamma[H\to f \bar f]&=&\Gamma[H\to f \bar f]_{\rm SM} (\xi_{H}^{f})^{2}\,,\\
\Gamma[H\to VV]&=&\Gamma[H\to V V]_{\rm SM}c_{\beta-\alpha}^{2}\,.
\eeqn
\eeqs
The next-to-leading-order (NLO) QCD corrections for the SM Higgs cases~\cite{Dittmaier:2011ti} are taken into account. The two-body and three-body partial decay widths of $H\to hh$ at the tree-level read~\cite{Djouadi:1995gv}
\beqs\label{eq:Hhhdecay}
\beqn
\Gamma[H\to hh]&=&\frac{\lambda_{Hhh}^{2}}{32\pi M_{H}}\Big( 1-4\kappa_{H}  \Big)^{1/2}\,,\\
\Gamma[H\to hh^{*}]&=&\frac{3 \lambda_{Hhh}^{2} m_{b}^{2}}{32\pi^{3} M_{H} v^{2}} (\xi_{H}^{d})^{2}\Big[ (\kappa_{H}-1)(2-\hf \log\kappa_{H})\non
&&+\frac{1-5\kappa_{H}}{\sqrt{4\kappa_{H}-1}}(\arctan\frac{2\kappa_{H}-1}{\sqrt{4\kappa_{H}-1}}-\arctan\frac{1}{\sqrt{4\kappa_{H}-1}}) \Big]\,,
\eeqn
\eeqs
with $\kappa_{H}\equiv M_{h}^{2}/M_{H}^{2}$. Two other decay modes of $H\to AZ$ and $H\to H^\pm W^\mp$ are also possible within the 2HDM. In order to highlight the $H\to hh$ decay mode in the mass range of $M_H\in (300\,,600)\,\GeV$, both decay modes will be kinematically suppressed with the large mass parameter inputs of $M_A=600\,\GeV$ and $M_{H^\pm}\gtrsim 600\,\GeV$.

\begin{figure}
\centering
\includegraphics[width=7cm,height=4.5cm]{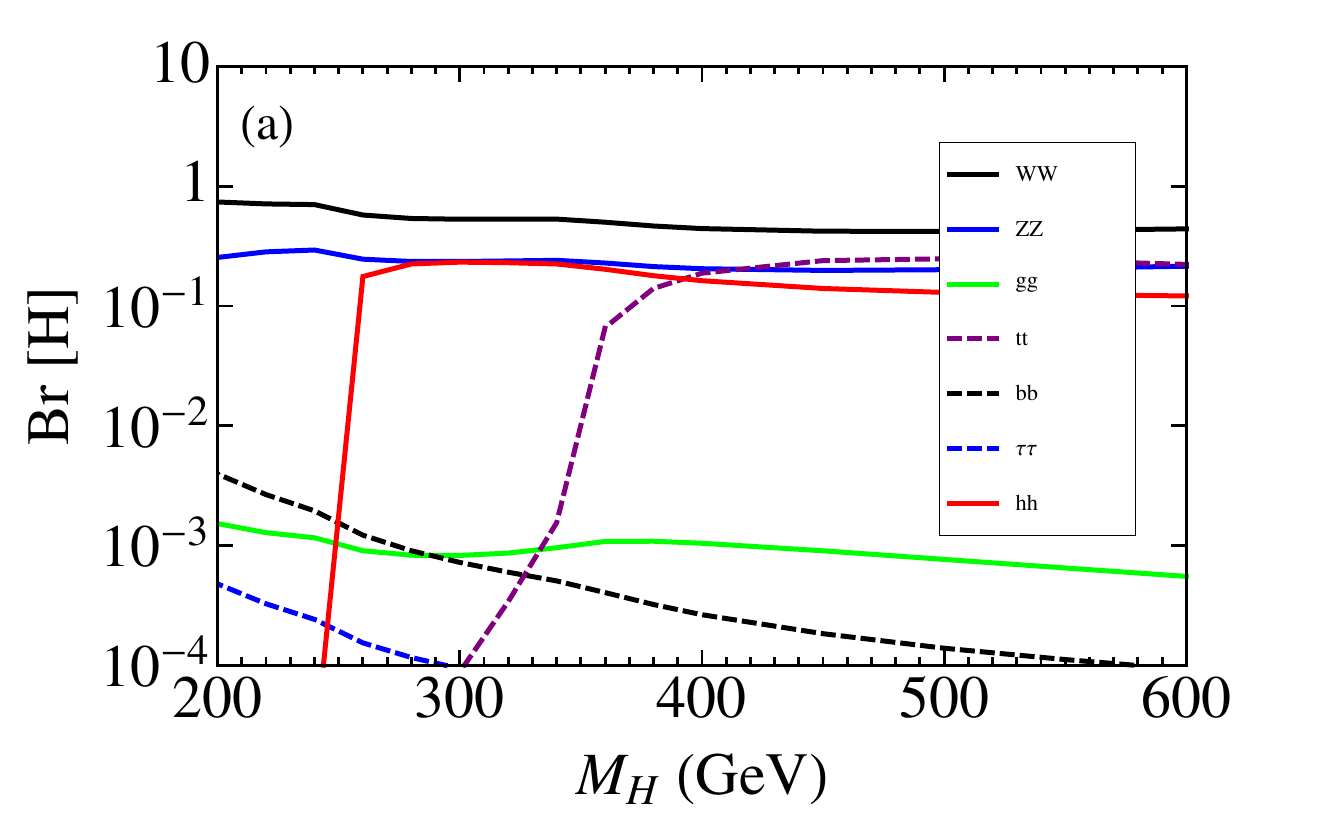}
\includegraphics[width=7cm,height=4.5cm]{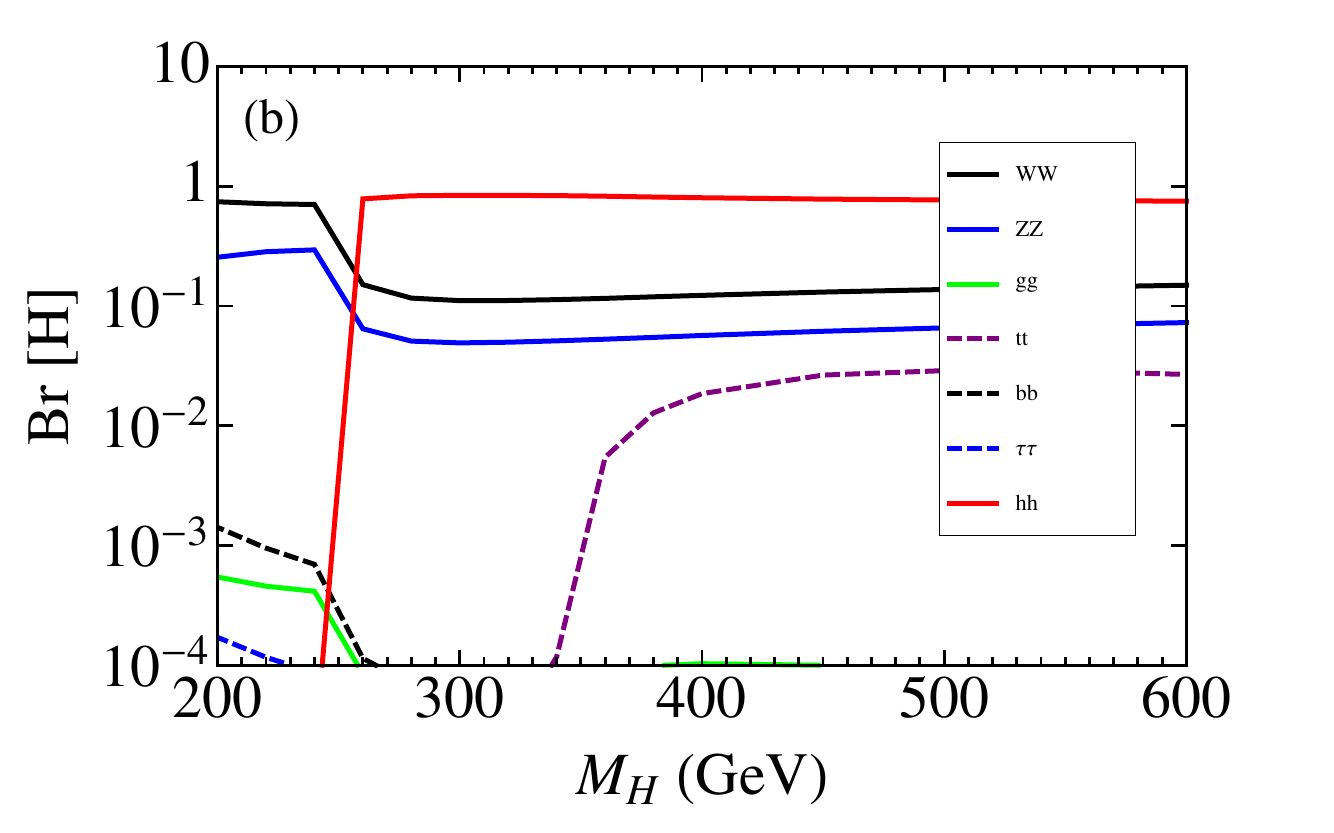}\\
\includegraphics[width=7cm,height=4.5cm]{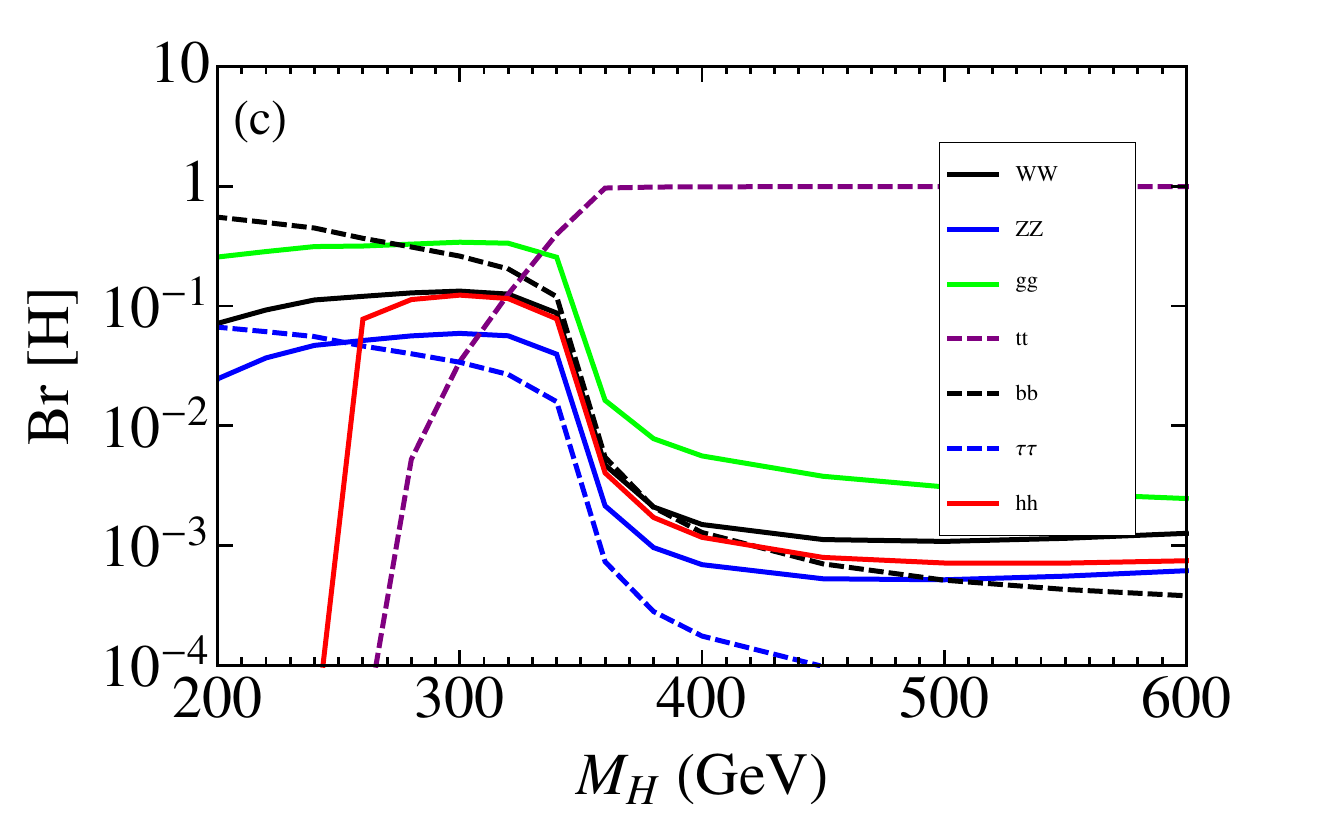}
\includegraphics[width=7cm,height=4.5cm]{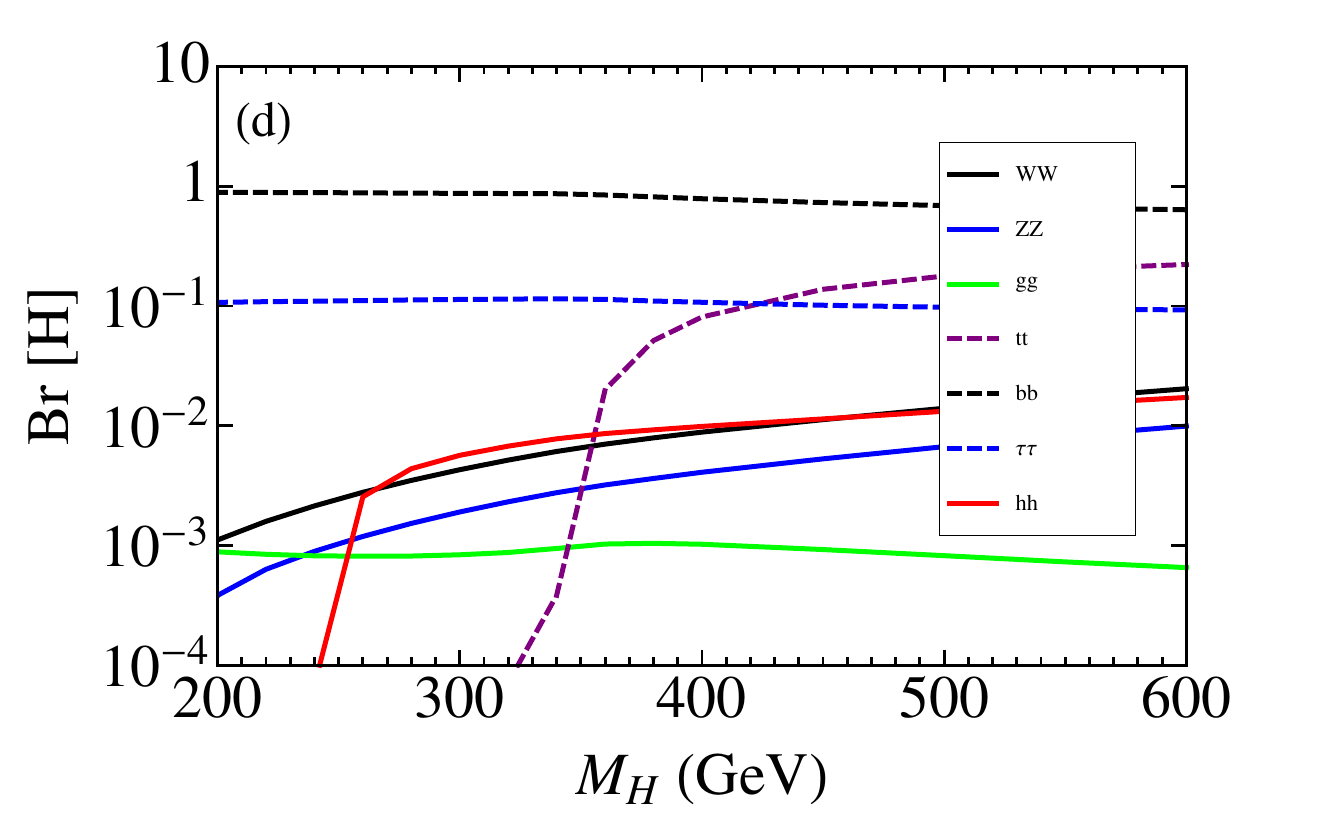}\\
\caption{\label{fig:HBR_mass} The decay branching ratios of $H$ in the mass range of $M_{H}\in (200\,,600)\,\GeV$. Upper left: 2HDM-I $(t_{\beta}=1)$. Upper right: 2HDM-I $(t_{\beta}=10)$. Lower left: 2HDM-II $(t_{\beta}=1)$. Lower right: 2HDM-II $(t_{\beta}=10)$. The alignment parameters follow Eq.~(\ref{eq:align_para}). }
\end{figure}

We also include the loop-induced decay modes of $H\to (gg\,,\gamma\gamma)$ by rescaling from the corresponding SM cases in the following manner:
\beqs\label{eq:Hdecay_SMloop}
\beqn
\Gamma[H\to gg]&=&\Gamma[H\to gg]_{\rm SM} \frac{\Big|\sum_{q=t\,,b}\xi_{H}^{q} A_{1/2}^{H}(\tau_{q}) \Big|^{2}}{\Big|A_{1/2}^{H}(\tau_{t}) \Big|^{2}}\,,\\
\Gamma[H\to \gamma\gamma]&=&\Gamma[H\to \gamma\gamma]_{\rm SM}\frac{\Big| \sum_{f} N_{c\,,f} Q_{f}^{2} \xi_{H}^{f} A_{1/2}^{H}(\tau_{f})+c_{\beta-\alpha}A_{1}^{H}(\tau_{W}) \Big|^{2}}{\Big|\sum_{f} N_{c\,,f} Q_{f}^{2} A_{1/2}^{H}(\tau_{f})+A_{1}^{H}(\tau_{W})  \Big|^{2}}\,,
\eeqn
\eeqs
with $\tau_{i}\equiv M_{H}^{2}/(4 M_{i}^{2})$. Here, $A_{1/2}^{H}(\tau)$ and $A_{1}^{H}(\tau)$ are the well-known form factors for the CP-even Higgs decaying into massless vector bosons through the spin-$1/2$ and spin-$1$ loops respectively. Semiquantitatively, the ratio for the most dominant partial decay widths of $H$ can be estimated as follows:
\beqn\label{eq:widRatio}
&&\Gamma[H\to VV]:\Gamma[H\to t \bar t]:\Gamma[H\to hh]\non
&\simeq& c_{\beta-\alpha}^{2} M_{H}^{4} : M_{H}^{2} m_{t}^{2} : c_{\beta-\alpha}^{2} (M_{A}^{2}+\lambda_{5} v^{2}+...)^{2}\,,
\eeqn
where numerical coefficients and phase space factors are neglected. One would envision the decay mode of $H\to hh$ to become dominant by tuning the inputs of $(M_{A}\,,\lambda_{5})$, even with the small alignment parameter input of $c_{\beta-\alpha}\sim\mO(10^{-1})-\mO(10^{-2})$. In Fig.~\ref{fig:Hhh_wid}, we demonstrate the partial decay width $\Gamma[H\to hh]$ for the 2HDM-I (left-panel) and 2HDM-II (right-panel) with the alignment parameter choices following Eq.~(\ref{eq:align_para}). For comparison, we also display the partial decay widths of $\Gamma[H\to hh]$ with the MSSM-like coupling term (\ref{eq:Hhh_MSSM}) for each case. Gathering all relevant partial decay widths in Eqs.~(\ref{eq:Hdecay_SMtree})-(\ref{eq:Hdecay_SMloop}), we display the ${\rm Br}[H]$ in the mass range of $M_{H}\in (200\,,600)\,\GeV$ in Fig.~\ref{fig:HBR_mass} with two different $t_{\beta}=1$ and $t_{\beta}=10$ inputs. Typically, a suppression of ${\rm Br}[H\to t \bar t]\sim \mO(10^{-1})-\mO(10^{-2})$ appears for the large $t_{\beta}=10$ input (right panels), due to that $\xi_{H}^{u}\sim 1/t_{\beta}$ along with the alignment limit for both 2HDM-I and 2HDM-II cases. Specifically, we have ${\rm Br}[H\to hh]\approx 1$ in the mass range of $250\,\GeV\lesssim M_{H}\lesssim 600\,\GeV$ for the 2HDM-I case with $t_{\beta}=10$. Thus, the experimental searches for the $H\to hh$ decay mode can be extended to a broad mass range instead of being restricted within the mass window of $250\,\GeV\lesssim M_{H}\lesssim 2m_{t}$, as was the case from the MSSM Higgs sector. The conventional experimental searches for the MSSM (2HDM-II type) CP-even Higgs were performed at the previous LEP experiments~\cite{Schael:2006cr} and the recent LHC $7\oplus 8\,\TeV$ runs~\cite{CMS:2013hja}. The LEP searches were made for the $HZ$ associated production and the Higgs pair productions, with the $H\to (b\bar b\,,\tau^+\tau^-)$ modes being mostly relevant for the searches. With the global fit to the 2HDM parameters, however, the $HZ$ associated production is highly suppressed. Therefore, the direct discovery potential of the heavy CP-even Higgs in the future $e^+e^-$ collider is challenging. The recent LHC search for $H$ was performed via the $\tau^+\tau^-$ final state, which is especially relevant for the large-$t_\beta$ inputs~\cite{CMS:2013hja}. For the input of $t_\beta=10$, the exclusion limit via the $H\to \tau^+\tau^-$ searches is about $M_H\gtrsim 400\,\GeV$ by naively assuming the mass degeneracy of $M_H\simeq M_A$ in the MSSM spectra.

\subsection{The productions and signals of $H$}

\begin{figure}
\centering
\includegraphics[width=6.8cm,height=4.5cm]{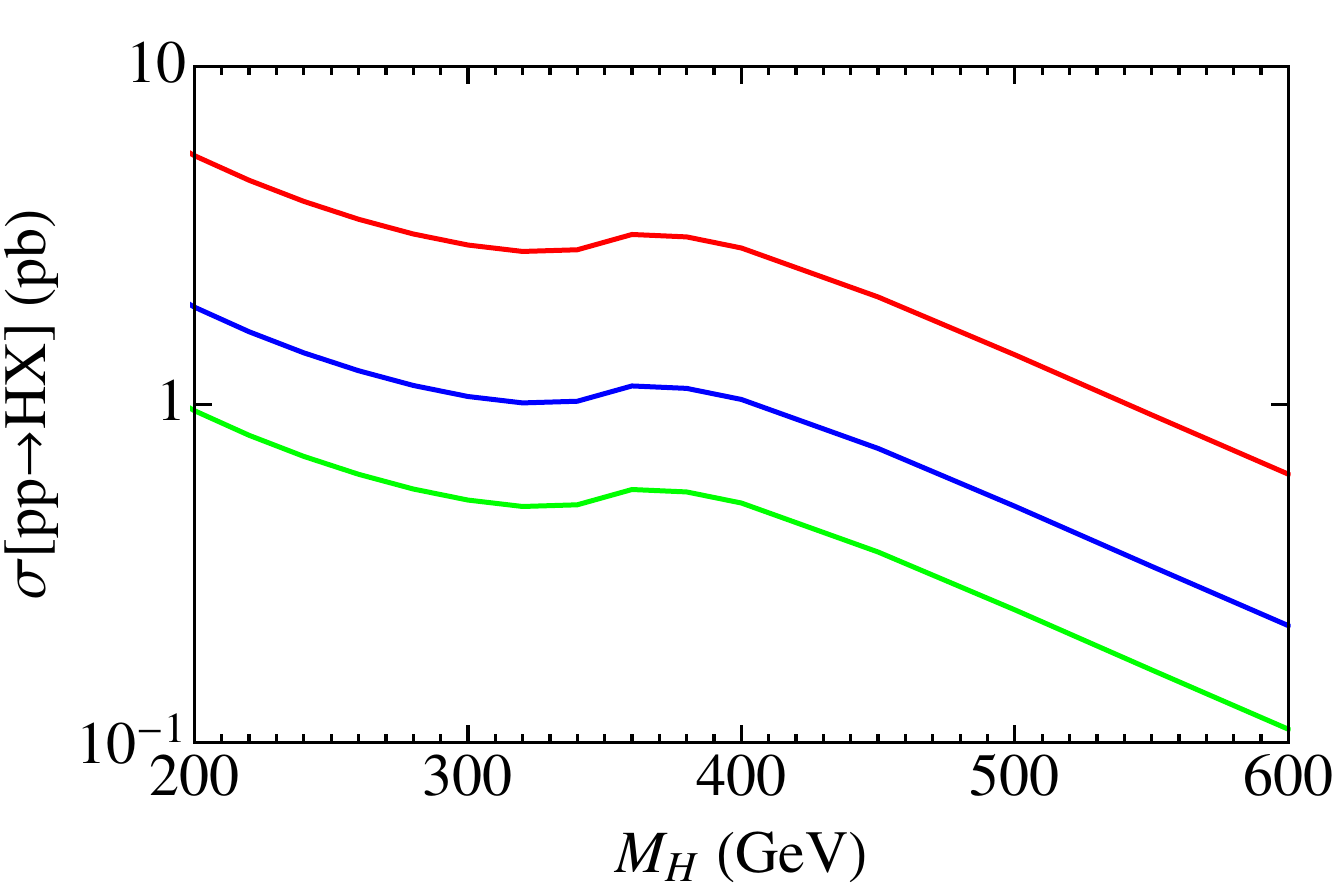}
\includegraphics[width=6.8cm,height=4.5cm]{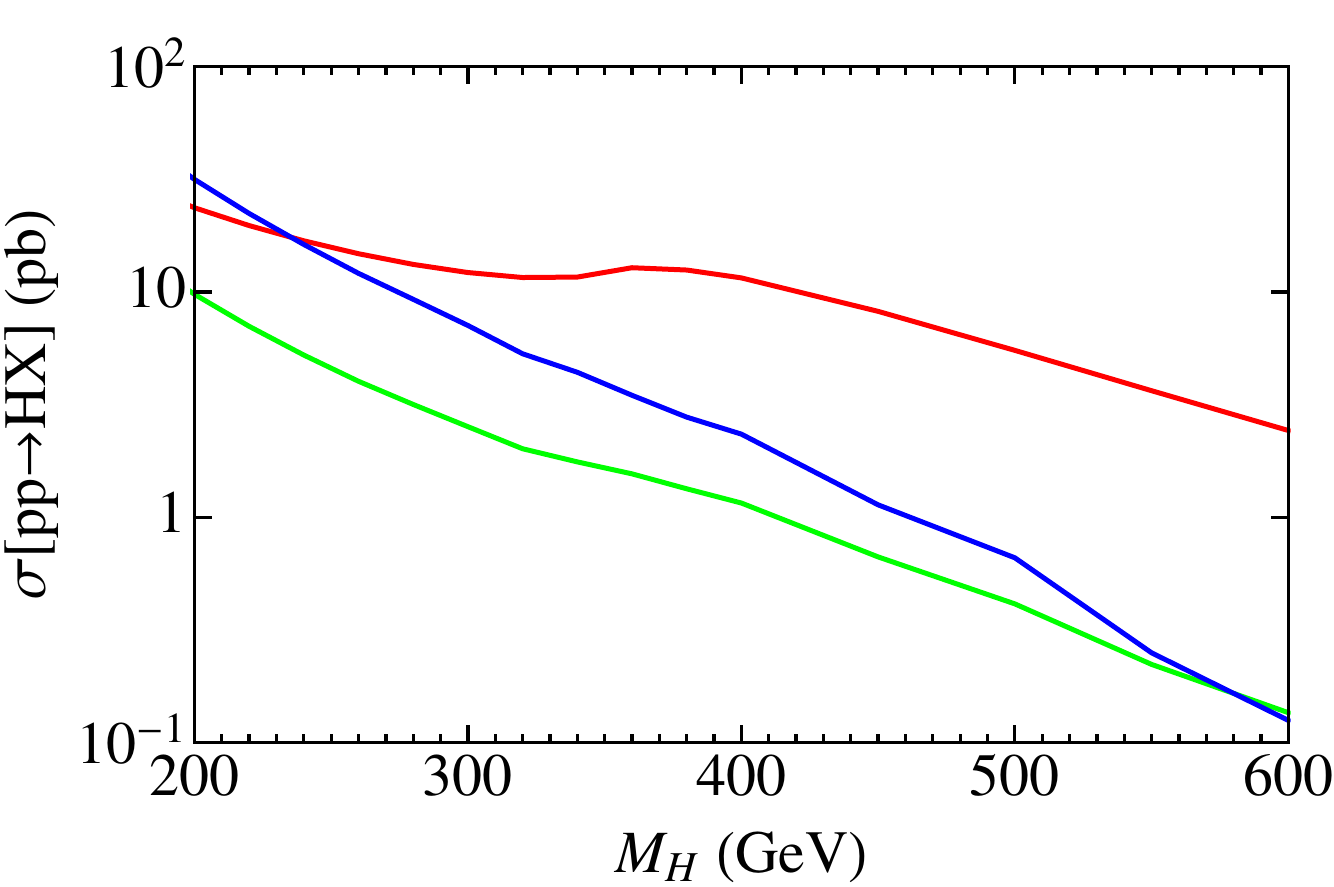}
\caption{\label{fig:Hprod_xsec} The inclusive heavy Higgs production cross sections $\sigma[pp\to HX]$ for $M_{H}\in (200\,,600)\,\GeV$: 2HDM-I (left) and 2HDM-II (right) at the LHC $14\,\TeV$ run. We show samples with $t_{\beta}=1$ (red), $t_{\beta}=5$ (green), and $t_{\beta}=10$ (blue) for each plot.}
\end{figure}

In general, $H$ can be produced via channels of (i) the gluon fusion, (ii) the vector boson fusion, (iii) associated productions with vector bosons, and (iv) associated productions with heavy quarks. The production cross sections for $H$ from channel (ii) and channel (iii) are highly suppressed by a factor of $\sim \mO(10^{-2})-\mO(10^{-4})$ to the corresponding SM cases when taking the alignment parameters in Eq.~(\ref{eq:align_para}). For the most dominant gluon fusion process, the cross sections are rescaled by
\beqn\label{eq:ggH_xsec}
\sigma[gg\to H]_{(\alpha\,,\beta)}&=&\sigma[gg\to H]_{\rm SM} \frac{\Gamma[H\to gg]_{(\alpha\,,\beta)}}{\Gamma[H\to gg]_{\rm SM}}\,,
\eeqn
where the NLO gluon fusion cross sections for the SM Higgs~\cite{Dittmaier:2011ti} will be used. The gluon fusion is the dominant production channel for most cases of our discussion; hence, the uncertainties of $\sim 10\%$ in the mass range of $M_H\in (300\,,600)\,\GeV$~\cite{Dittmaier:2011ti} roughly set the uncertainties for the signal evaluations. Likewise, for the b-quark associated production channels, the relevant processes involve $(0\,,1\,,2)$ b-quarks in the final states:
\beqn\label{eq:bbH_process}
&&b \bar b\to H\,,~~ b/\bar b g\to b/\bar b H\,,~~ q \bar q/gg\to H b\bar b\,.
\eeqn
We also include these cross sections by rescaling from the corresponding SM cases:
\beqn\label{eq:bbH_xsec}
&&\frac{~~\sigma[b\bar b\to H]_{(\alpha\,,\beta)}}{\sigma[b\bar b\to H]_{\rm SM}}=\frac{~~\sigma[b/\bar b g\to b/\bar b H]_{(\alpha\,,\beta)}}{\sigma[b/\bar b g\to b/\bar b H]_{\rm SM}}=\frac{~~\sigma[pp\to b\bar b H]_{(\alpha\,,\beta)}}{\sigma[pp\to b\bar b H]_{\rm SM}}=(\xi_{H}^{d})^{2}\,.
\eeqn
These b-quark associated cross sections~\cite{Campbell:2002zm, Harlander:2003ai, Dittmaier:2003ej} are given by including the NLO QCD corrections. Given that $\xi_{H}^{d}\to t_{\beta}$ along with the alignment limit for the 2HDM-II, it is apparent that the corresponding inclusive cross sections associated with $b$ quarks in Eq.~(\ref{eq:bbH_xsec}) would become sizable.

\begin{figure}
\centering
\includegraphics[width=6.8cm,height=4.5cm]{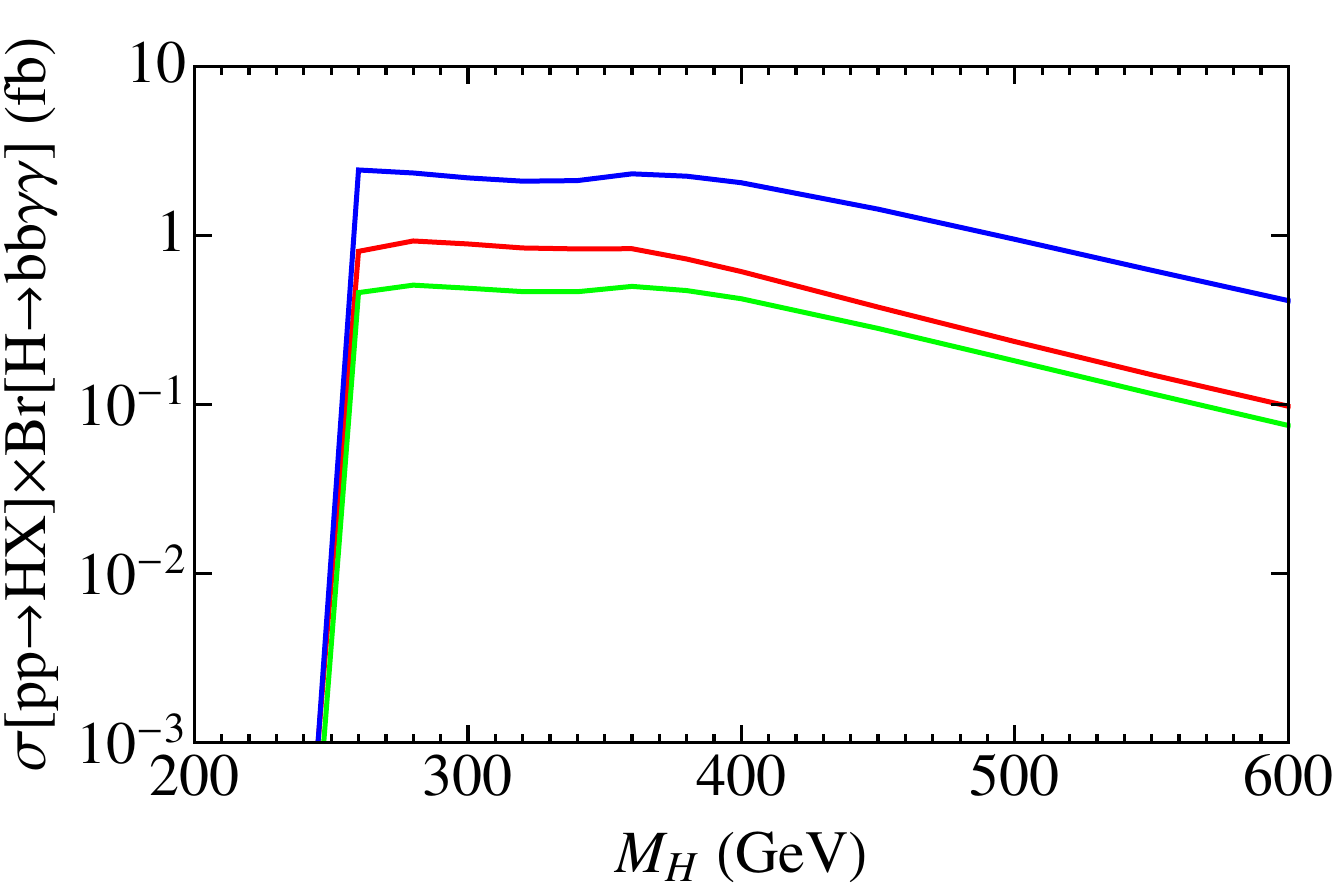}
\includegraphics[width=6.8cm,height=4.5cm]{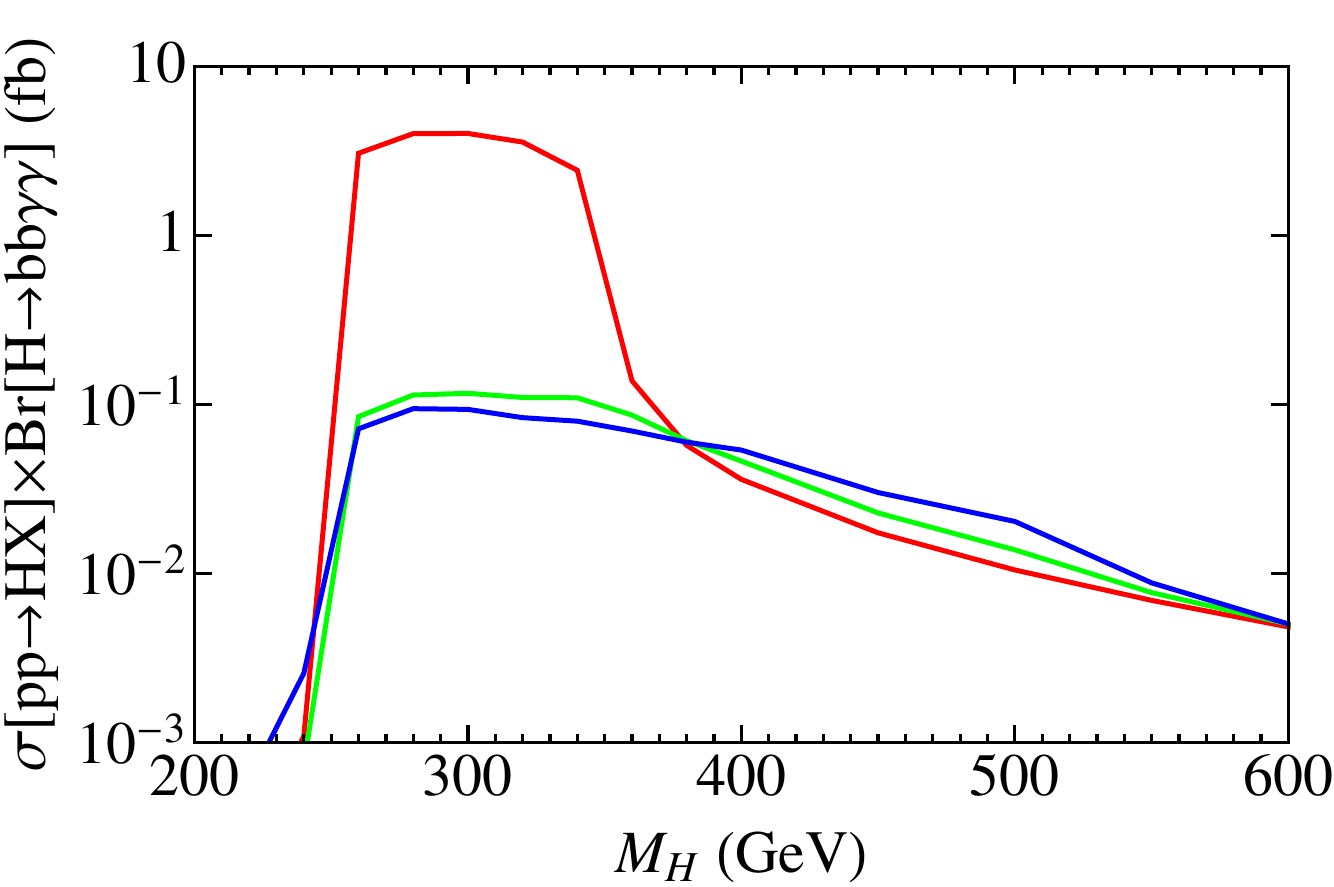}
\caption{\label{fig:Hbbgaga_sig} The $\sigma[pp\to HX]\times {\rm Br}[H\to b \bar b \gamma\gamma]$ for $M_{H}\in (200\,,600)\,\GeV$: 2HDM-I (left) and 2HDM-II (right) at the LHC $14\,\TeV$ run. We show samples with $t_{\beta}=1$ (red), $t_{\beta}=5$ (green), and $t_{\beta}=10$ (blue) for each plot.}
\end{figure}

\begin{figure}
\centering
\includegraphics[width=6.8cm,height=4.5cm]{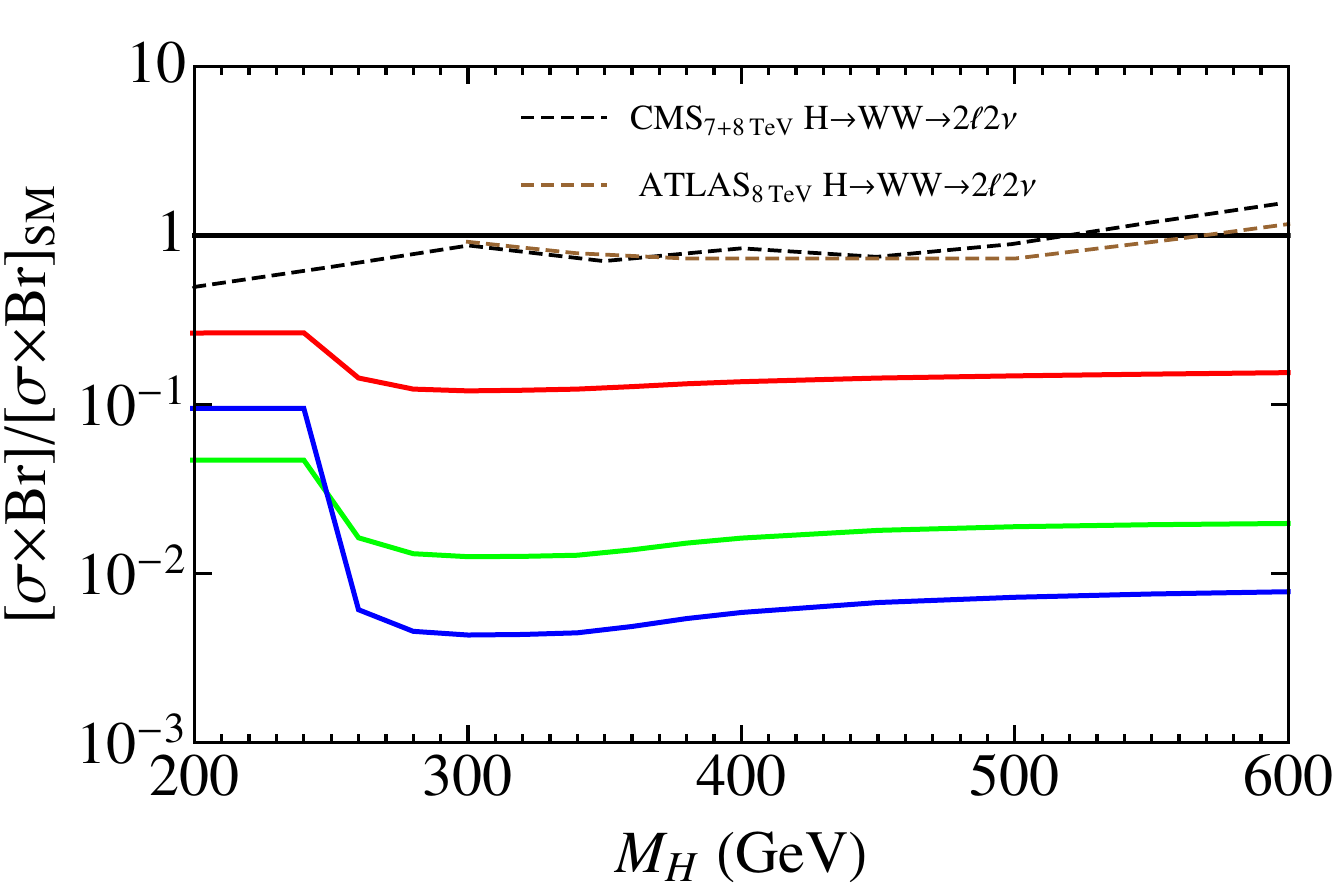}
\includegraphics[width=6.8cm,height=4.5cm]{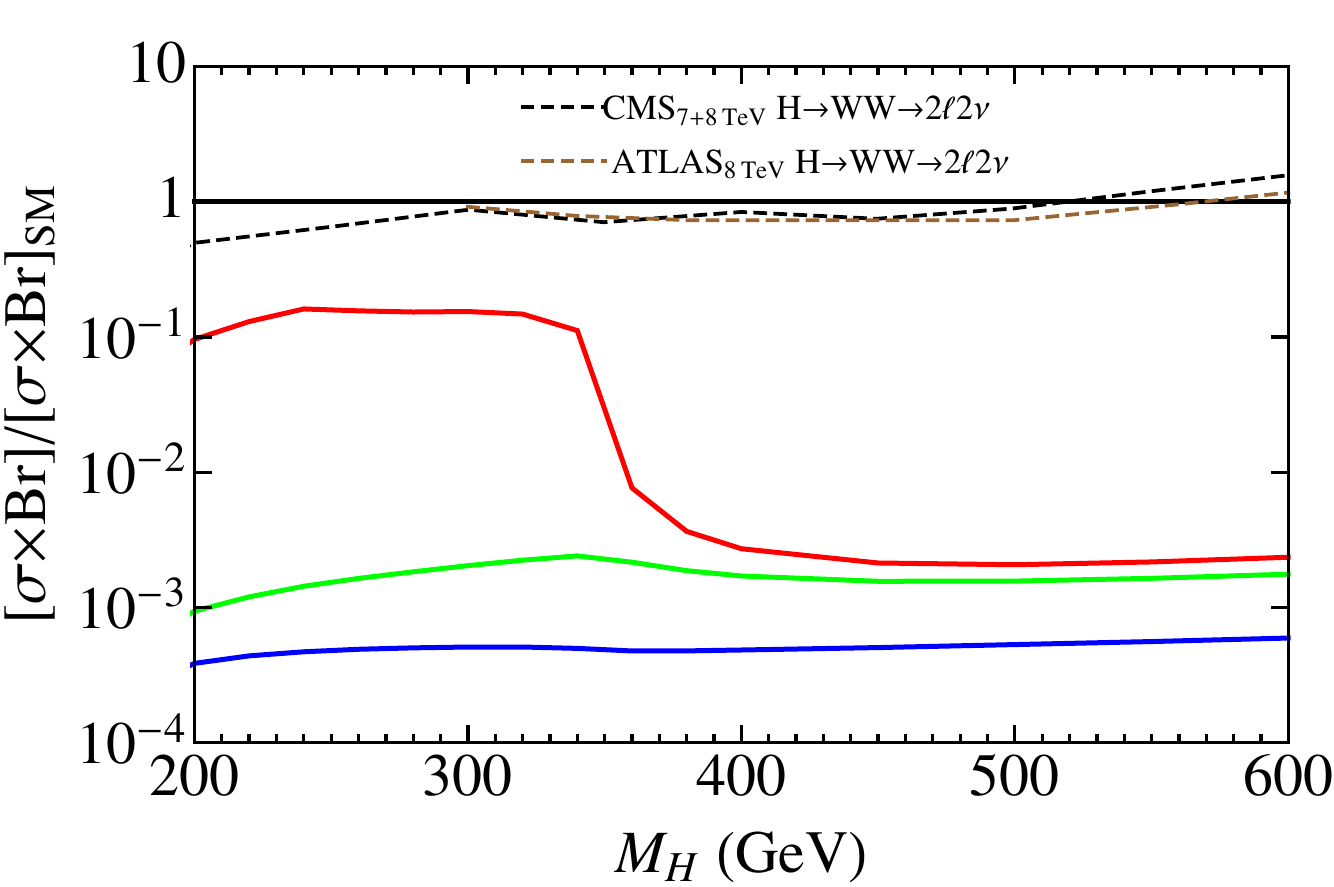}\\
\includegraphics[width=6.8cm,height=4.5cm]{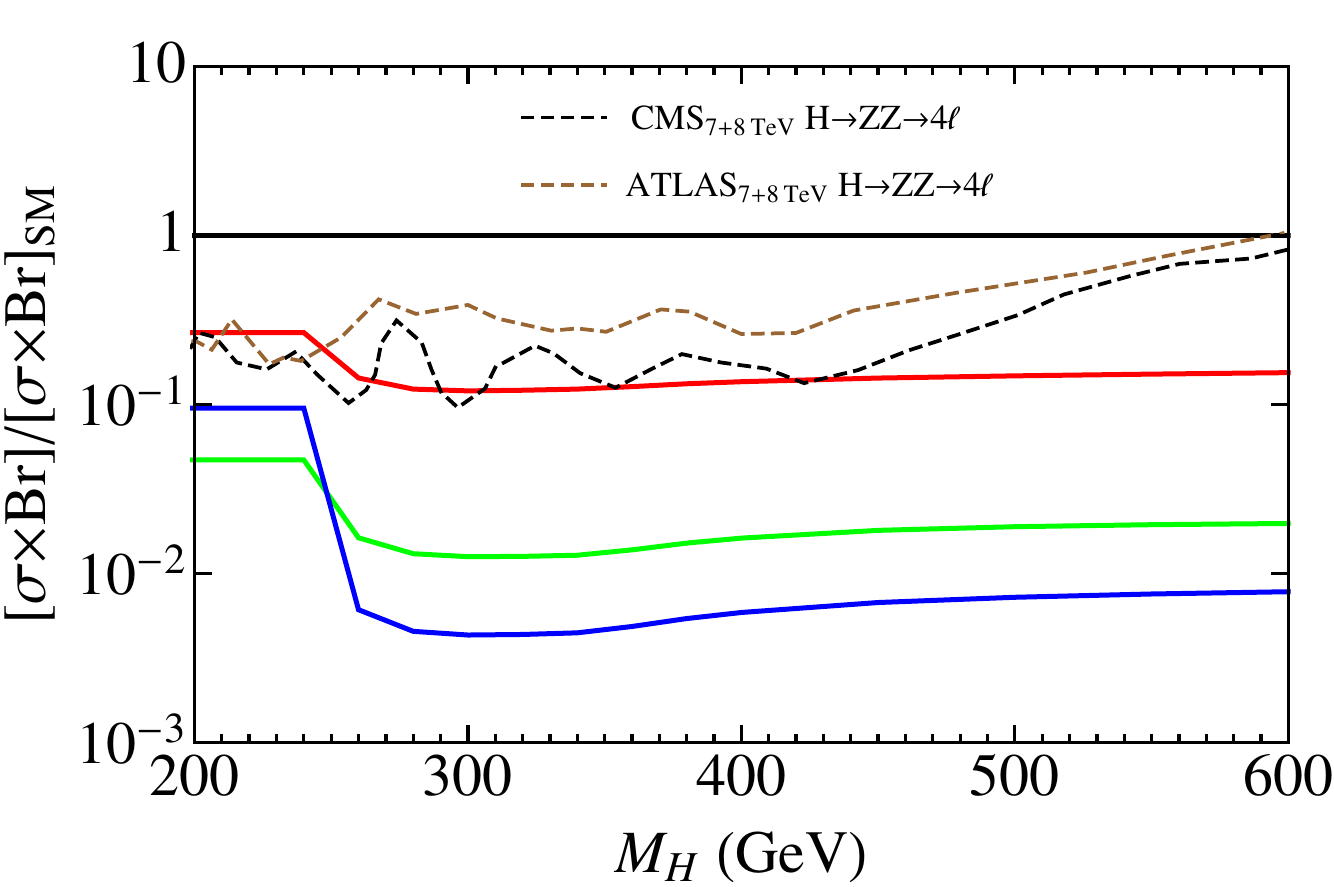}
\includegraphics[width=6.8cm,height=4.5cm]{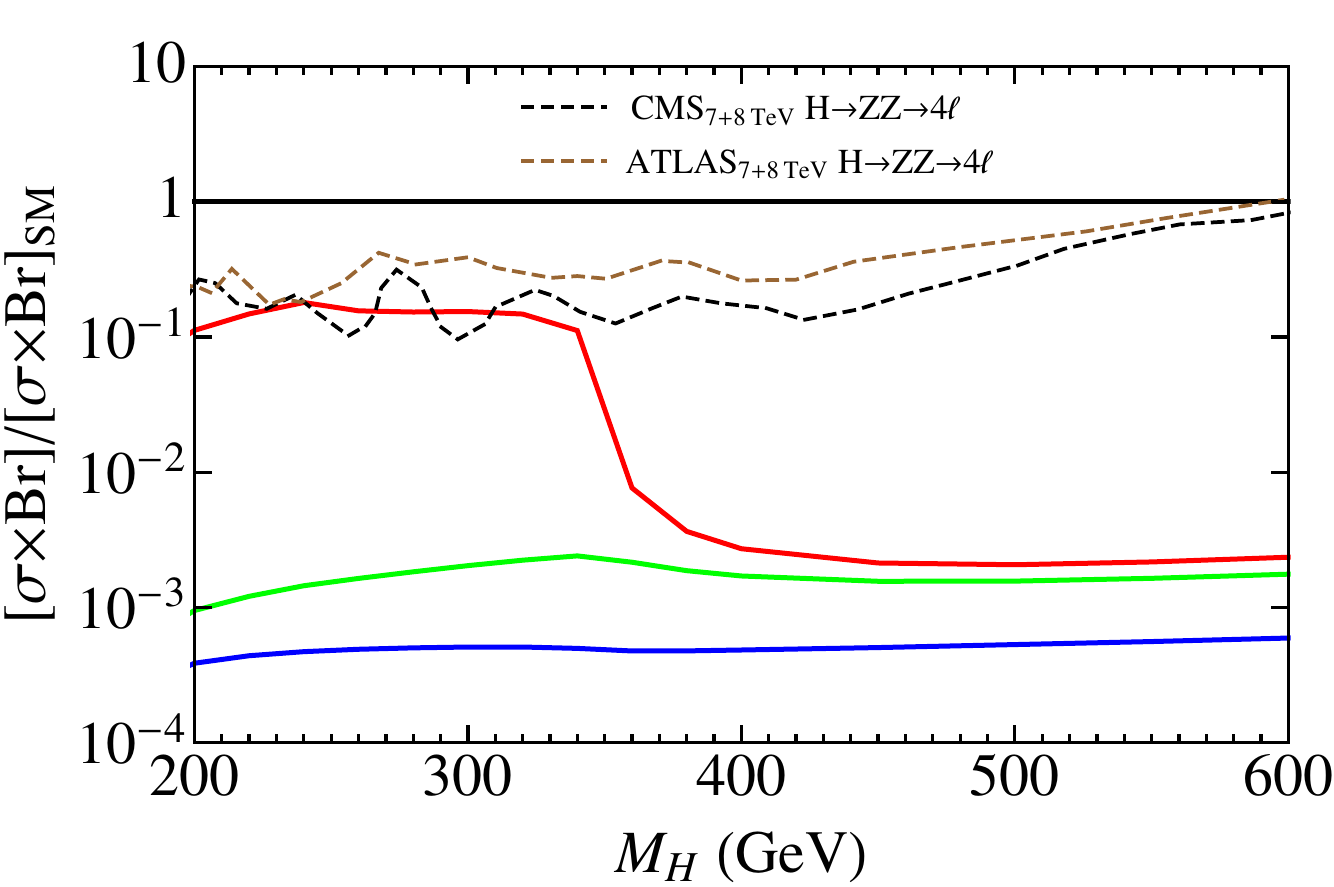}\\
\caption{\label{fig:H_excl} The direct experimental bounds on $H$ via the $H\to WW\to 2\ell 2\nu$ and $H\to ZZ\to 4\ell$ channels for $M_{H}\in (200\,,600)\,\GeV$. We demonstrated the exclusions for $H$ in both 2HDM-I (left panels) and 2HDM-II (right-panels) cases. The alignment parameter inputs of $c_{\beta-\alpha}$ follow Eq.~(\ref{eq:align_para}), and inputs of $t_{\beta}=1$ (red), $t_{\beta}=5$ (green), and $t_{\beta}=10$ (blue) are shown for each plot.}
\end{figure}

The inclusive production cross sections of $H$ are shown in Fig.~\ref{fig:Hprod_xsec} for the mass range of $M_{H}\in (200\,,600)\,\GeV$. For the 2HDM-I case, essentially only the gluon fusion process is accounted for while the $b$-quark associated processes are negligible. However, for the 2HDM-II case, the $b$-quark associated productions are enhanced by the large-$t_{\beta}$ inputs since $\xi_{H}^{d}\to t_{\beta}$ along with the alignment limit. Combining the production cross sections and the decay branching ratios evaluated previously, we display the $\sigma[pp\to HX]\times {\rm Br}[H\to b \bar b \gamma\gamma]$ in Fig.~\ref{fig:Hbbgaga_sig} for both 2HDM-I and 2HDM-II. Without performing the detailed kinematic analysis, the $\sigma[pp\to HX]\times {\rm Br}[H\to b \bar b \gamma\gamma]$ results are roughly promising for the upcoming LHC runs at $14\,\TeV$ with the integrated luminosities accumulated up to $\int \mL dt\sim \mO(10^{2})-\mO(10^{3})\,\fb^{-1}$.

By the end of this section, we will briefly mention the current experimental searches for $H$ via the conventional $H\to WW\to 2\ell 2\nu$ and $H\to ZZ\to 4\ell$ channels. In Fig.~\ref{fig:H_excl}, we display the signal predictions of $H$ via these leptonic channels versus the experimental sensitivities reached from both ATLAS~\cite{ATLAS:2012bmv, TheATLAScollaboration:2013zha} and CMS~\cite{CMS:yxa, CMS:eya} by the $7\oplus 8\,\TeV$ data. Because of the freedom of setting the $(M_A\,,\lambda_5)$ inputs, together with the alignment parameter inputs (\ref{eq:align_para}), the sensitivities to these conventional modes can be suppressed by $\mO(10^{-2})-\mO(10^{-3})$ compared with the current LHC searches for the large $t_\beta$ cases. Consequently, the alternative mode of $H\to hh$ can be considered prior to the conventional $(WW\,,ZZ)$ final state searches for the heavy-mass $H$.


\section{The Analysis of The $H\to hh\to b \bar b \gamma\gamma$ Signals}
\label{section:2b2gamma}

In this section, we analyze the LHC searches for $H$ via the $b\bar b\gamma\gamma$ final states, which can be potentially promising for the enhanced cubic scalar coupling $\lambda_{Hhh}$ cases. The dominant SM background processes include $b \bar b\gamma\gamma$ and $t \bar t\gamma\gamma$, while other contributions from $ h(\to\gamma\gamma)Z(\to b \bar b)$, $h(\to \gamma\gamma) t\bar t$, and $h(\to b\bar b)h(\to \gamma\gamma)$ are negligible~\cite{Barger:2013jfa}. For two leading irreducible background processes, potential contributions from the reducible QCD backgrounds with jets to fake either $b$ jets and/or photons are not negligible. These fake rates for the relevant QCD background processes were considered in earlier studies on the SM Higgs self-coupling probes~\cite{Baur:2003gp}. Here we follow the ATLAS detector performance~\cite{Aad:2009wy} in the photon identification efficiencies:
\beqn\label{eq:photonfake}
&&\epsilon_{q\to \gamma}\approx 3.6\times 10^{-4} \,,~~ \epsilon_{g\to \gamma}\approx 3.6\times 10^{-5}\,,
\eeqn
with $q$ and $g$ representing the quark jet and gluon jet, respectively. The $b$-jet mistag rates are taken to be
\beqn\label{eq:bfake}
&&\epsilon_{c\to b}\approx 0.2\,,~~ \epsilon_{j\to b}\approx 0.01\,,
\eeqn
with $j$ representing the light jets, i.e., jets which are neither $b$ tagged or $c$ tagged. For the $b\bar b\gamma\gamma$ background, the relevant reducible QCD background contributions include
\beqn\label{eq:bbaa_reduc}
&& c\bar c \gamma\gamma\,,~~ j j \gamma\gamma\,,~~ b\bar b g \gamma\,, ~~ c \bar c g\gamma\,,~~ j j g \gamma\,,~~ b\bar b q \gamma\,,~~ c \bar c q\gamma\,,~~ j j q \gamma\,,\non
&& b\bar b g g\,,~~ c \bar c g g\,,~~ j j g g\,, ~~ b\bar b g q\,,~~ c \bar c g q\,,~~ j j g q\,, ~~ b\bar b q q\,,~~ c \bar c q q\,,~~ j j q q\,;
\eeqn
while for the $t \bar t\gamma\gamma$ background, the relevant reducible background contributions include
\beqn\label{eq:ttaa_reduc}
&& t \bar t g \gamma\,,~~ t \bar t q \gamma\,,~~ t \bar t g g\,,~~ t \bar t g q\,,~~ t \bar t q q\,.
\eeqn
All contributions from the reducible QCD background are estimated by the cross section of each process listed in Eqs.~(\ref{eq:bbaa_reduc}) and (\ref{eq:ttaa_reduc}), as weighted by the $b$-jet and photon mis-identification rates of Eqs.~(\ref{eq:photonfake}) and (\ref{eq:bfake}). According to our evaluation, these reducible backgrounds contribute to $\sigma_{\rm reduc}(b\bar b\gamma\gamma)/\sigma_{\rm tot}(b\bar b\gamma\gamma)\approx 25\%$ and $\sigma_{\rm reduc}(t\bar t\gamma\gamma)/\sigma_{\rm tot}(t\bar t\gamma\gamma)\approx 4\%$ of the total background cross sections, respectively. By including the uncertainties due to the NLO QCD corrections and the parton distribution functions, we estimated the uncertainties of the background processes to be $\sim 8\,\%$. In the kinematic analysis below, we always assume the same cut efficiencies between the irreducible background processes and the corresponding reducible background processes.

To generate events for the signal processes, we obtain a Universal FeynRules Output~\cite{Degrande:2011ua} simplified model containing $H$ as the only BSM particle. The necessary coupling terms to be implemented include: the cubic $Hhh$ coupling, the dimension-five $Hgg$ and $h\gamma\gamma$ couplings, and the $H(h) b \bar b$ Yukawa couplings. We generate events at the parton level for both signal and background processes by {\sc Madgraph}/{\sc MadEvent}~\cite{Alwall:2011uj}. Afterwards, the results are passed to {\sc PYTHIA}~\cite{Sjostrand:2006za} for simulating the initial- and final-state radiation, parton showering, and hadronization. Eventually, we pass all events to {\sc Delphes}~\cite{deFavereau:2013fsa} for the fast detector simulation, where we use the default ATLAS detector card. For the jet clustering, we adopt the anti-$k_{T}$ jet algorithm with the parameter $R=0.6$ for the ATLAS detector card. In addition, we take an overall $b$-tagging efficiency of $70\,\%$ for all of the kinematic regions.

\subsection{Optimization of kinematic cuts}

To identify the signals from the background, we start with the preliminary cuts by selecting the events with $b$ jets and photons:
\beqn\label{eq:bbaa_selection}
&&n_{b}\ge2 \,,~~ n_{\gamma}=2\,.
\eeqn
These photons and $b$ jets should also satisfy the cuts on their pseudorapidities, the transverse momenta, and the mutual $\eta-\phi$ distances:
\beqn\label{eq:bbaa_PT_Eta}
&&|\eta_{\gamma\,,b}|<2.5\,,~~ p_{T\,,\gamma}>25\,\GeV\,,~~ p_{T\,,b}>25\,\GeV\,,\non
&&\Delta R(b\,,b)>0.4\,,~~ \Delta R(\gamma\,,\gamma)>0.4\,,~~ \Delta R(b\,,\gamma)>0.4\,.
\eeqn
To further reduce the $t\bar t\gamma\gamma$ background~\cite{Baglio:2012np}, we veto events containing leptons with the transverse momenta of $p_{T\,,\ell}>20\,\GeV$ and the pseudorapidities of $|\eta_{\ell}|<2.5$.

\begin{figure}
\centering
\includegraphics[width=7.5cm,height=4.5cm]{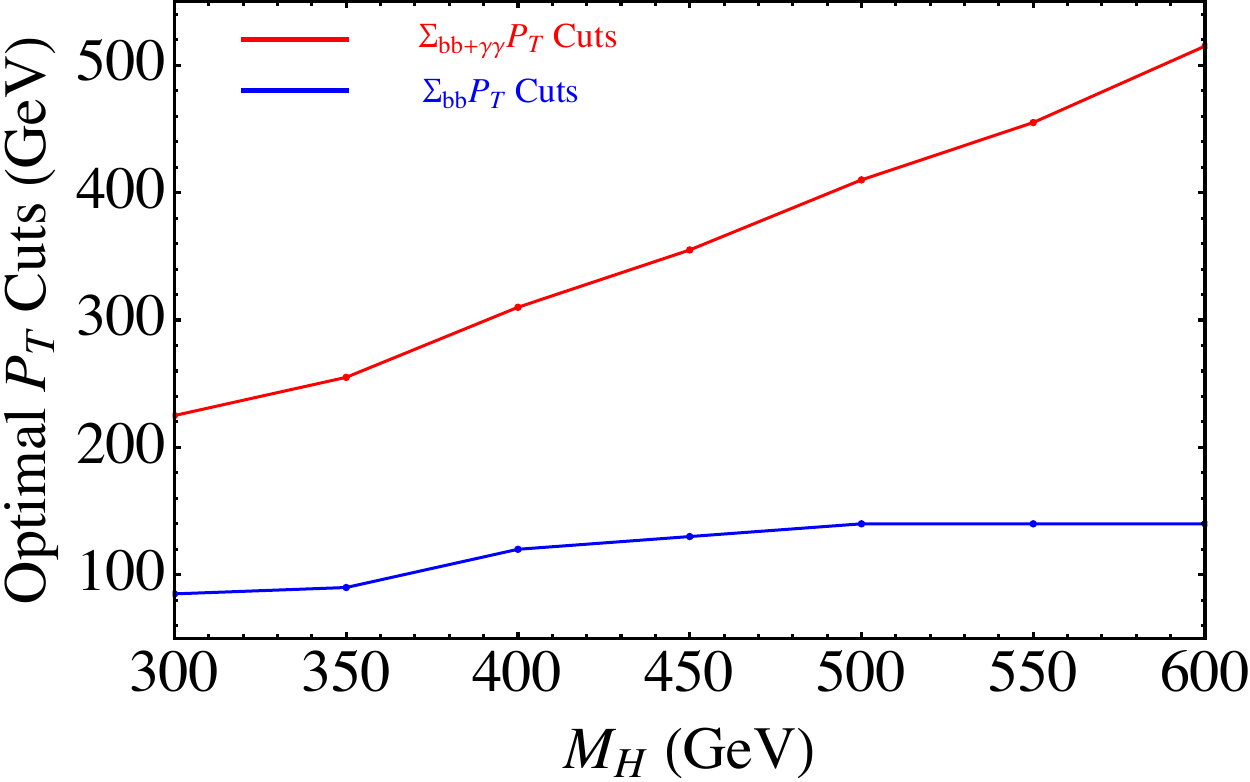}
\caption{\label{fig:PTopt_mass} The most optimal cuts on $\sum_{b} p_{T}$ and $\sum_{b\,,\gamma} p_{T}$ for $H$ in the mass range of $M_{H}\in (300\,,600)\,\GeV$.}
\end{figure}

For larger $M_{H}$ inputs, it is generally efficient to select events containing hard $b$ jets and photons. Hence, we impose cuts on the sum of transverse momenta of the selected $b$ jets and photons, which are expected to increase with larger $M_{H}$ inputs. In practice, we scan over the cuts on the $p_{T}$ summations for both heavy Higgs boson signals and the SM background processes in the range of $\sum_{b} p_{T}\in (50\,,300)\,\GeV$ and $\sum_{b\,,\gamma} p_{T}\in (100\,,600)\,\GeV$, respectively. Afterwards, the most optimal cuts on $\sum_{b}p_{T}$ and $\sum_{b\,,\gamma} p_{T}$ for each $M_{H}$ input are selected by those yielding the largest $S/B$. The mass dependence of the most optimal cuts of $(\sum_{b} p_{T}\,, \sum_{b\,,\gamma} p_{T})$ on the $M_{H}$ inputs are displayed in Fig.~\ref{fig:PTopt_mass}.

\begin{figure}
\centering
\includegraphics[width=6.8cm,height=6cm]{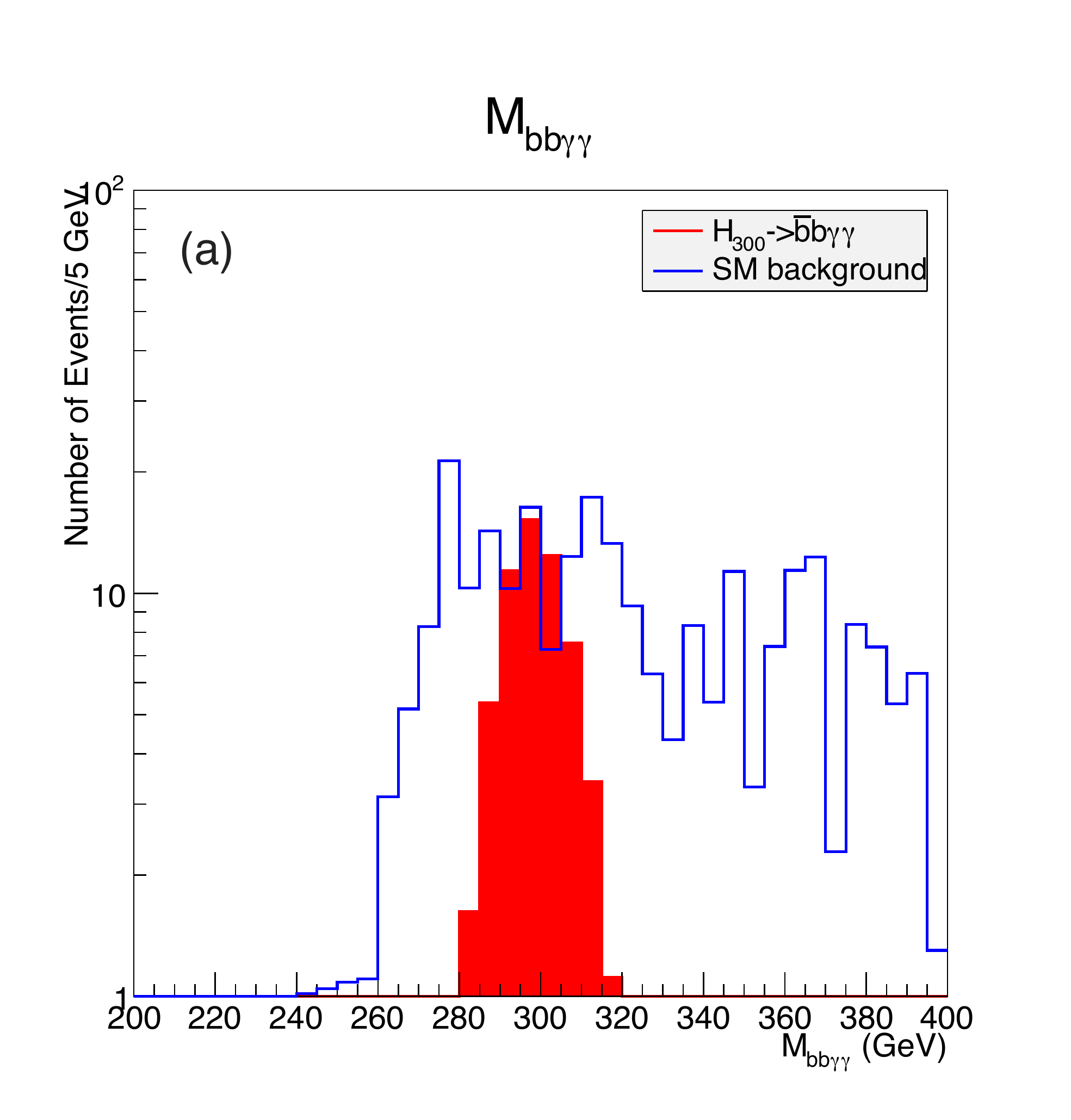}
\includegraphics[width=6.8cm,height=6cm]{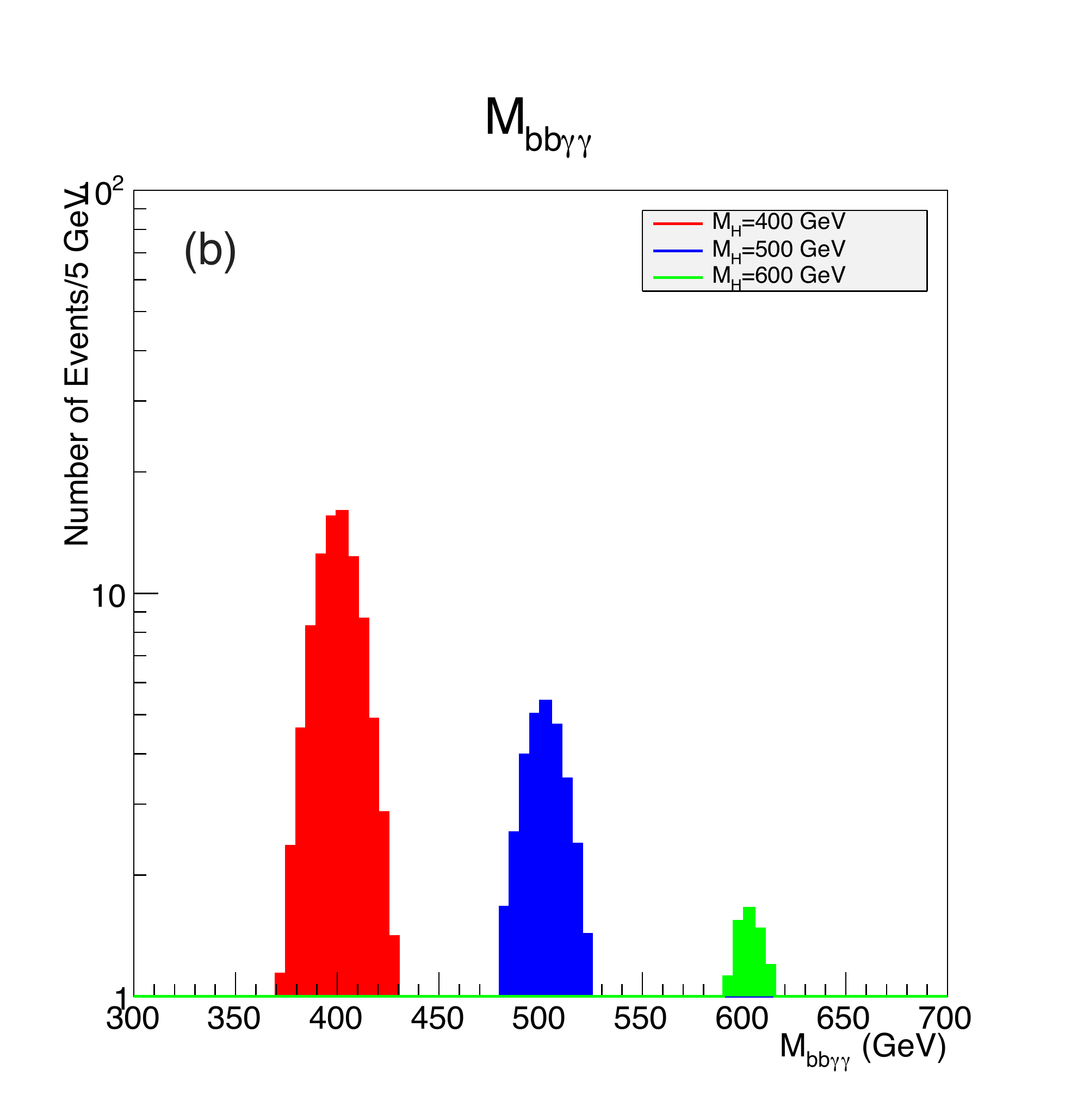}
\caption{\label{fig:mbbaa} Left: The $m_{bb\gamma\gamma}$ distributions for both signal process with $M_{H}=300\,\GeV$ (red) and background (blue) processes. The nominal cross section of $\sigma[pp\to HX]\times {\rm Br}[H\to b\bar b\gamma\gamma]$ is taken to be $1\,\fb$. Right: The $m_{bb\gamma\gamma}$ distributions for signal processes with $M_{H}=(400\,,500\,,600)\,\GeV$. The corresponding signal cross sections are taken for the 2HDM-I with the $t_{\beta}=10$ input. Both plots are evaluated for the LHC $14\,\TeV$ run with $\int \mL dt=1000\,\fb^{-1}$. }
\end{figure}

It is also straightforward to impose the invariant mass cuts on the selected two $b$ jets and two photons around the mass window of $125\,\GeV$ light CP-even Higgs~\cite{Baglio:2012np}:
\beqn\label{eq:invm_cuts}
&&112.5\,\GeV< m_{bb}< 137.5\,\GeV\,,~~ 120\,\GeV< m_{\gamma\gamma}<130\,\GeV\,,
\eeqn
where the mass resolutions for photons and $b$ jets are taken into account~\cite{ATLAS:2011xda, TheATLAScollaboration:2013lia}. For events containing more than two $b$ jets, we pair all possible combinations and find the one with $m_{bb}$ mostly close to $125\,\GeV$. After imposing the selection conditions listed above, the invariant mass of $m_{bb\gamma\gamma}$ should reconstruct the mass window for the particular $M_{H}$ input. In Fig.~\ref{fig:mbbaa}, the invariant mass distributions of $m_{bb\gamma\gamma}$ are shown after imposing the event selection cuts (\ref{eq:bbaa_selection}), the optimal $p_{T}$ summation cuts given in Fig.~\ref{fig:PTopt_mass}, and the $(m_{bb}\,,m_{\gamma\gamma})$ cuts (\ref{eq:invm_cuts}) sequentially. For the $M_{H}=300\,\GeV$ case, we demonstrated the distributions for both signal and background processes, with the corresponding optimal $p_{T}$ summation cuts imposed for this sample. On the right panel of Fig.~\ref{fig:mbbaa}, the $m_{bb\gamma\gamma}$ distributions are shown for the signals with the $M_{H}=(400\,,500\,,600)\,\GeV$ cases, with the signal cross sections set by the $t_{\beta}=10$ input along with the alignment limit of the 2HDM-I. By observing the $m_{bb\gamma\gamma}$ distributions for various $M_{H}$ inputs, we require the mass window cuts of $m_{bb\gamma\gamma}$ to be the following:
\beqn\label{eq:bbaa_window}
M_{H}=300\,\GeV&:& m_{bb\gamma\gamma}\in (275\,,335)\,\GeV\,,\non
M_{H}=350\,\GeV&:& m_{bb\gamma\gamma}\in (295\,,405)\,\GeV\,,\non
M_{H}=400\,\GeV&:& m_{bb\gamma\gamma}\in (355\,,450)\,\GeV\,,\non
M_{H}=450\,\GeV&:& m_{bb\gamma\gamma}\in (400\,,510)\,\GeV\,,\non
M_{H}=500\,\GeV&:& m_{bb\gamma\gamma}\in (455\,,560)\,\GeV\,,\non
M_{H}=550\,\GeV&:& m_{bb\gamma\gamma}\in (500\,,615)\,\GeV\,,\non
M_{H}=600\,\GeV&:& m_{bb\gamma\gamma}\in (555\,,665)\,\GeV\,.
\eeqn

\begin{table}[t]
\centering
\begin{tabular}{c||c|c|c|c|c}
\hline
 Cuts &  $\sigma_{\rm total}$ & $b\bar b\gamma\gamma$ selection  & $p_{T}$ sum  & $(m_{bb}\,,m_{\gamma\gamma})$ & $m_{bb\gamma\gamma}$   \\\hline\hline  
 $H(300\,\GeV)\to b \bar b\gamma\gamma\,({\rm fb})$ & $1$ & $0.17$  & $0.13$ & $0.06$ &  $0.06$ \\
 $b \bar b \gamma\gamma\,({\rm fb})$ & $6.729\times 10^{3}$ & $98.8$  & $44.4$ &  $0.28$ & $0.12$   \\
 $t \bar t \gamma\gamma\,({\rm fb})$ & $11.5$ & $0.51$ & $0.46$ & $2.4\times 10^{-3}$  & $7.5\times 10^{-4}$ \\
 $S/\sqrt{B}$  & $0.06$ & $0.42$ & $0.54$ & $3.58$ & $5.48$  \\
\hline
 \end{tabular}
\caption{The event cut efficiency for the LHC $14\,\TeV$ run of the signal and background processes, where both irreducible and reducible background processes are taken into account. We impose the cuts of $b \bar b\gamma\gamma$ selections (\ref{eq:bbaa_selection}), the optimal $p_{T}$ sum cuts as given in Fig.~\ref{fig:PTopt_mass}, the invariant mass cuts of (\ref{eq:invm_cuts}) and (\ref{eq:bbaa_window}). The signal reaches are estimated for the integrated luminosity of $\int\mL dt=1000\,\fb^{-1}$, where we also included the uncertainties of the SM background processes. }\label{tab:eff_300}
\end{table}

With the most optimal kinematic cuts at hand, we impose them sequentially to both signal and background processes. The results are given in Table.~\ref{tab:eff_300} for the $M_{H}=300\,\GeV$ sample, with a nominal cross section of $\sigma[pp\to HX]\times{\rm Br}[H\to b \bar b\gamma\gamma]=1\,\fb$ taken for the signal process. The evaluation of the cut efficiencies $S/\sqrt{B}$ are performed for the $\int \mL dt=1000\,\fb^{-1}$ case, where we also took the systematic uncertainties into account. As for the other $M_H$ inputs, similar kinematic cuts will be imposed to evaluate the efficiencies for the background suppression.

\subsection{Implications to the LHC searches for $H$ in the general 2HDM}

\begin{figure}
\centering
\includegraphics[width=6.3cm,height=4.5cm]{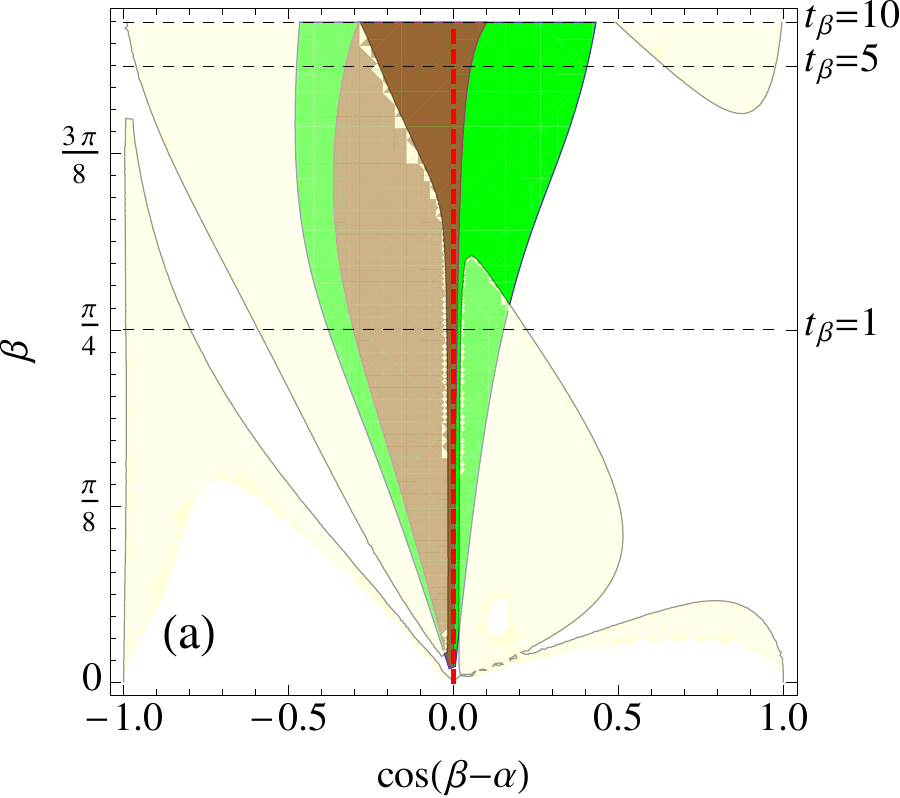}
\includegraphics[width=6.3cm,height=4.5cm]{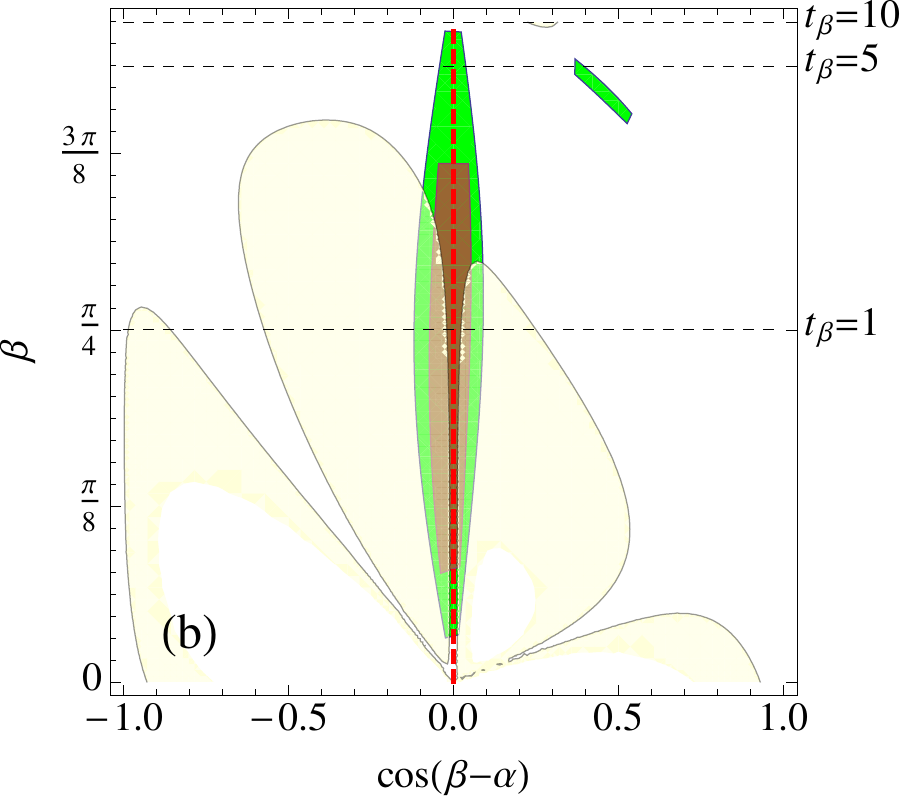}\\
\includegraphics[width=6.3cm,height=4.5cm]{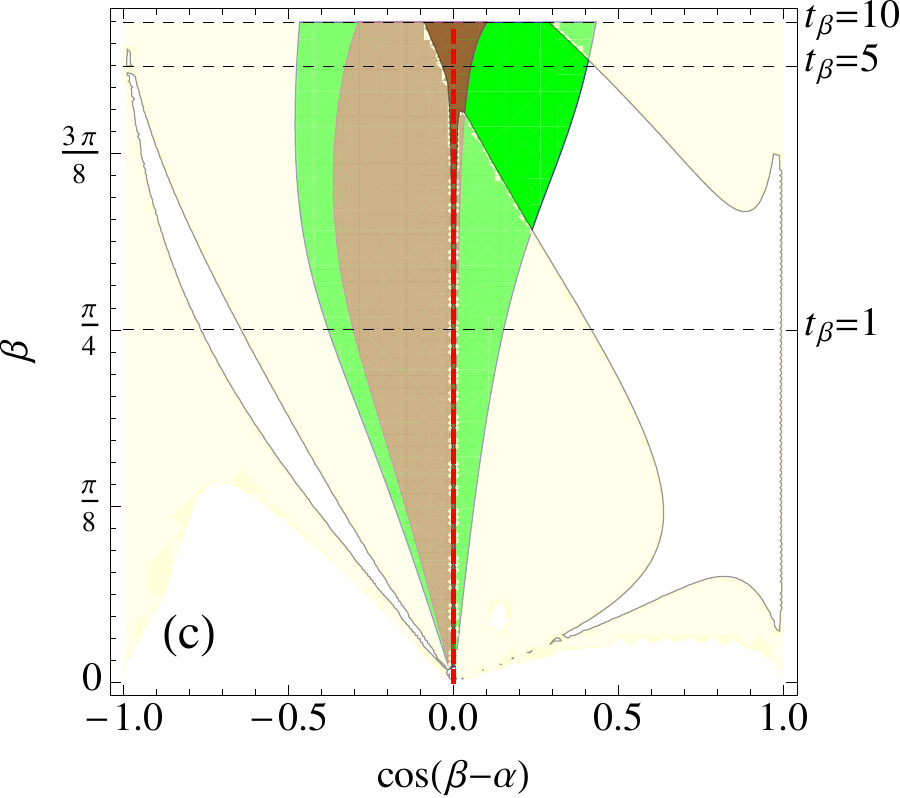}
\includegraphics[width=6.3cm,height=4.5cm]{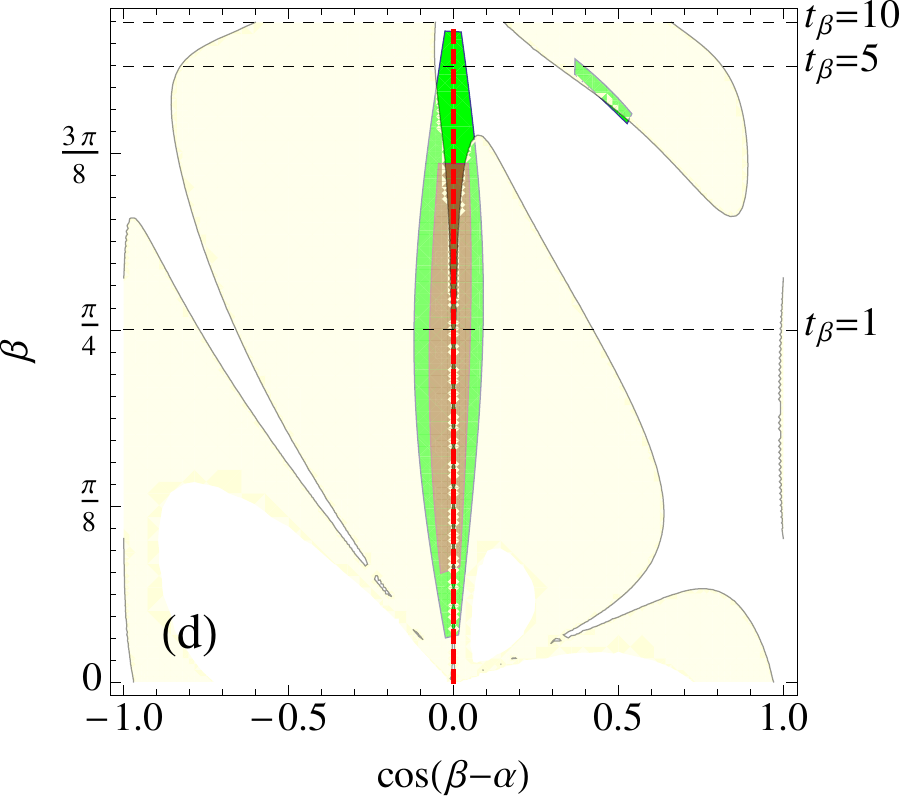}\\
\caption{\label{fig:H300_search} The LHC 14 search sensitivities of the $H\to b\bar b\gamma\gamma$ final states on the $(c_{\beta-\alpha}\,,\beta)$ parameter space. Upper left: 2HDM-I for $\int \mL dt=100\,\fb^{-1}$. Upper right: 2HDM-II for $\int \mL dt=100\,\fb^{-1}$. Lower left: 2HDM-I for the $\int \mL dt=3000\,\fb^{-1}$. Lower right: 2HDM-II for $\int \mL dt=3000\,\fb^{-1}$. The yellow shadow in each plot represents the parameter regions within the reach via the $b\bar b\gamma\gamma$ final states. The green and brown bends are the global fit to the $125\,\GeV$ Higgs boson in the 2HDM at the $68\,\%$ and $95\,\%$ C.L.}
\end{figure}

\begin{figure}
\centering
\includegraphics[width=6.8cm,height=4.5cm]{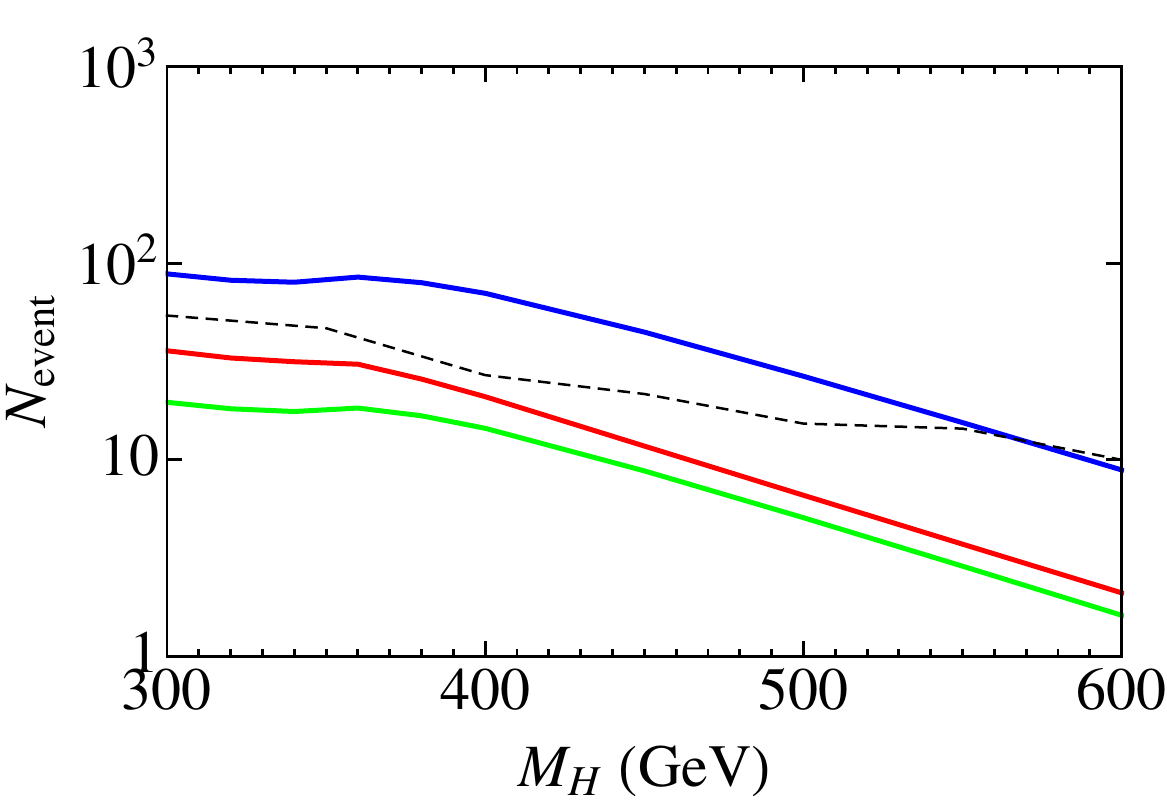}
\includegraphics[width=6.8cm,height=4.5cm]{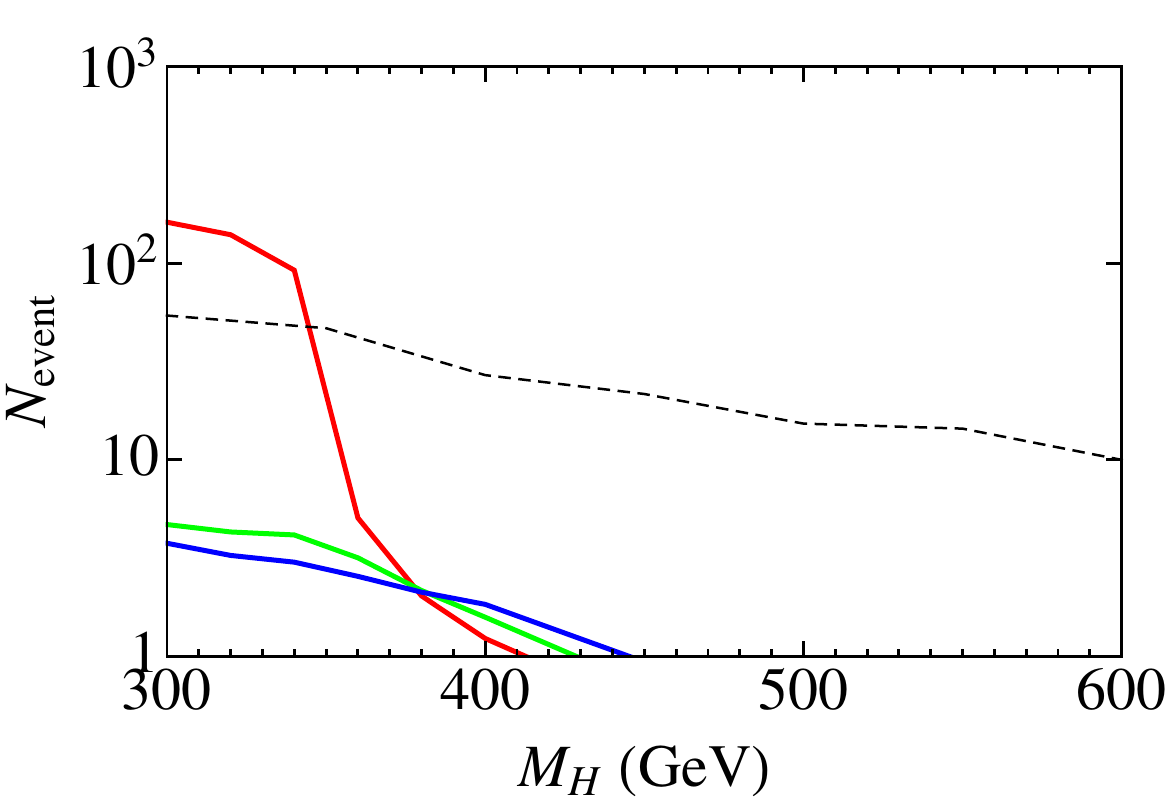}\\
\includegraphics[width=6.8cm,height=4.5cm]{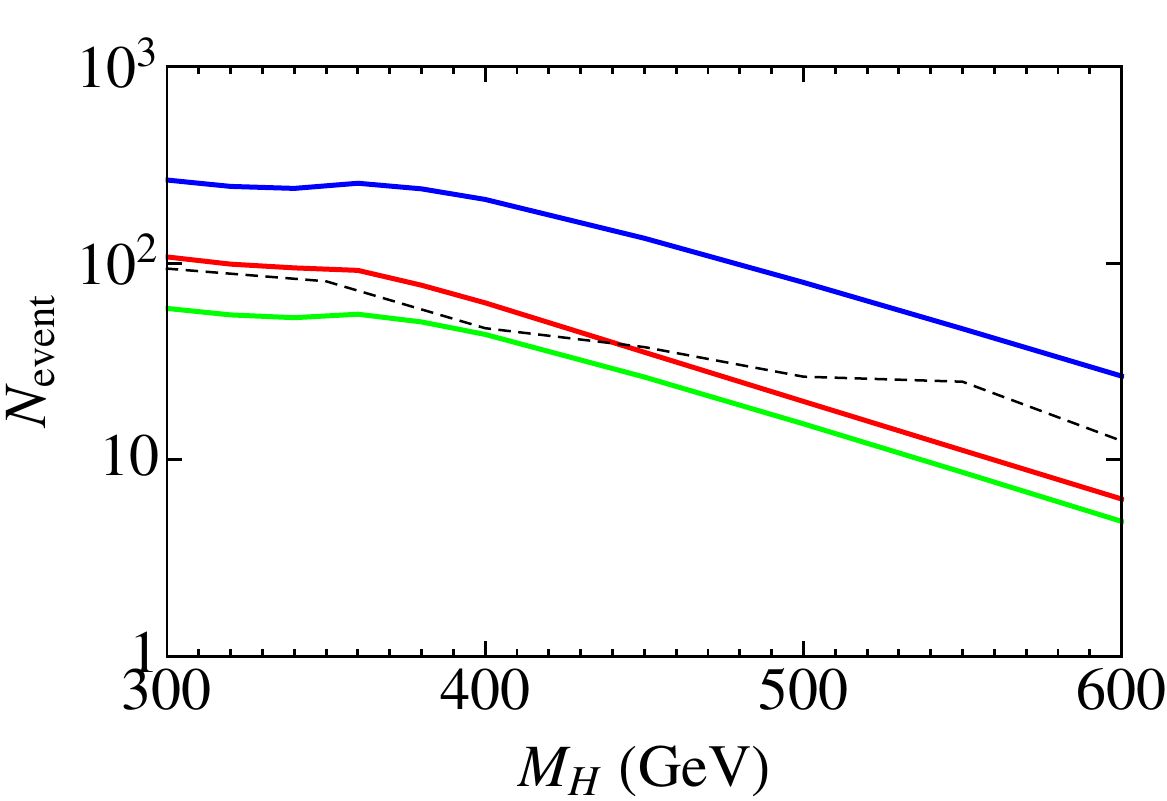}
\includegraphics[width=6.8cm,height=4.5cm]{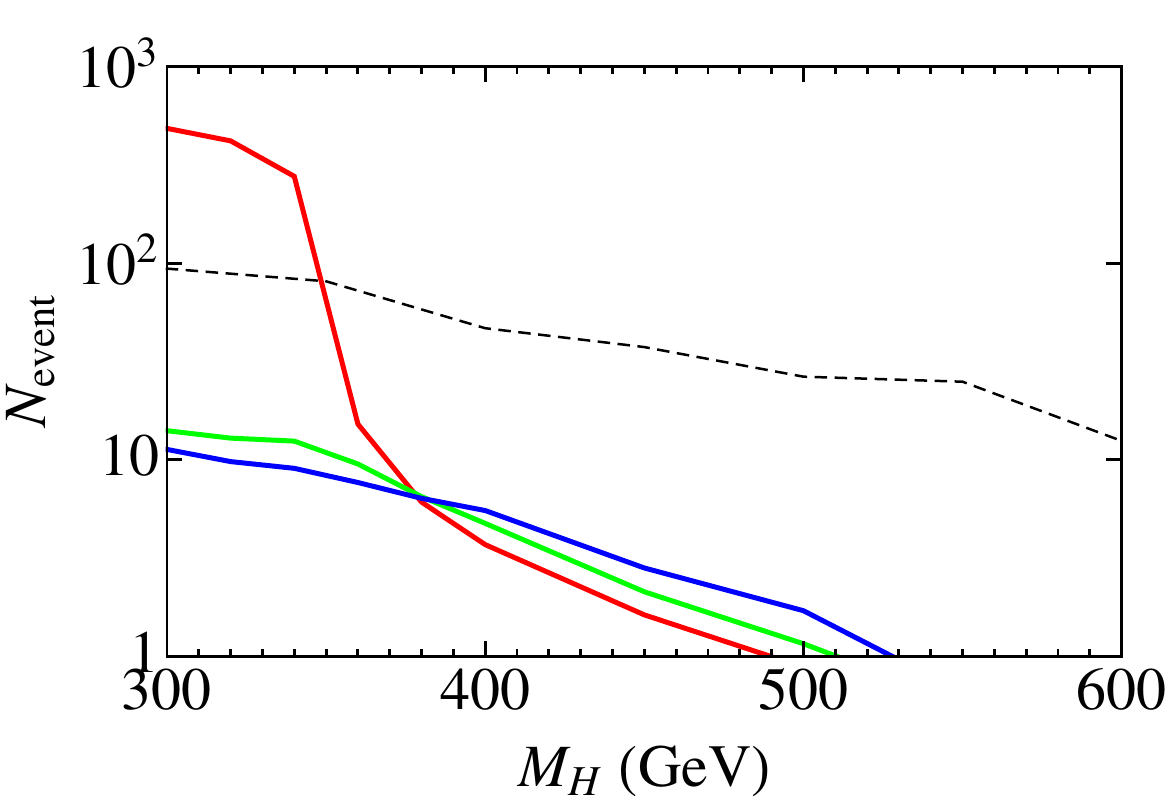}\\
\caption{\label{fig:Hmass_reach} The number of events of the $H\to b\bar b\gamma\gamma$ final states in contrast to the background contributions. Upper left: 2HDM-I for $\int \mL dt=1000\,\fb^{-1}$. Upper right: 2HDM-II for $\int \mL dt=1000\,\fb^{-1}$. Lower left: 2HDM-I for $\int \mL dt=3000\,\fb^{-1}$. Lower right: 2HDM-II for $\int \mL dt=3000\,\fb^{-1}$. We show samples with $t_{\beta}=1$ (red), $t_{\beta}=5$ (green), and $t_{\beta}=10$ (blue) for each plot, while the alignment parameter choices of Eq.~(\ref{eq:align_para}) are followed. The discovery limit (black dashed curve) of ${\rm max}\{5\sqrt{B}\,,10  \}$ is demonstrated for each sample with the $\int \mL dt=1000\,\fb^{-1}$ and $\int \mL dt=3000\,\fb^{-1}$ cases.}
\end{figure}

\begin{figure}
\centering
\includegraphics[width=6.8cm,height=4.5cm]{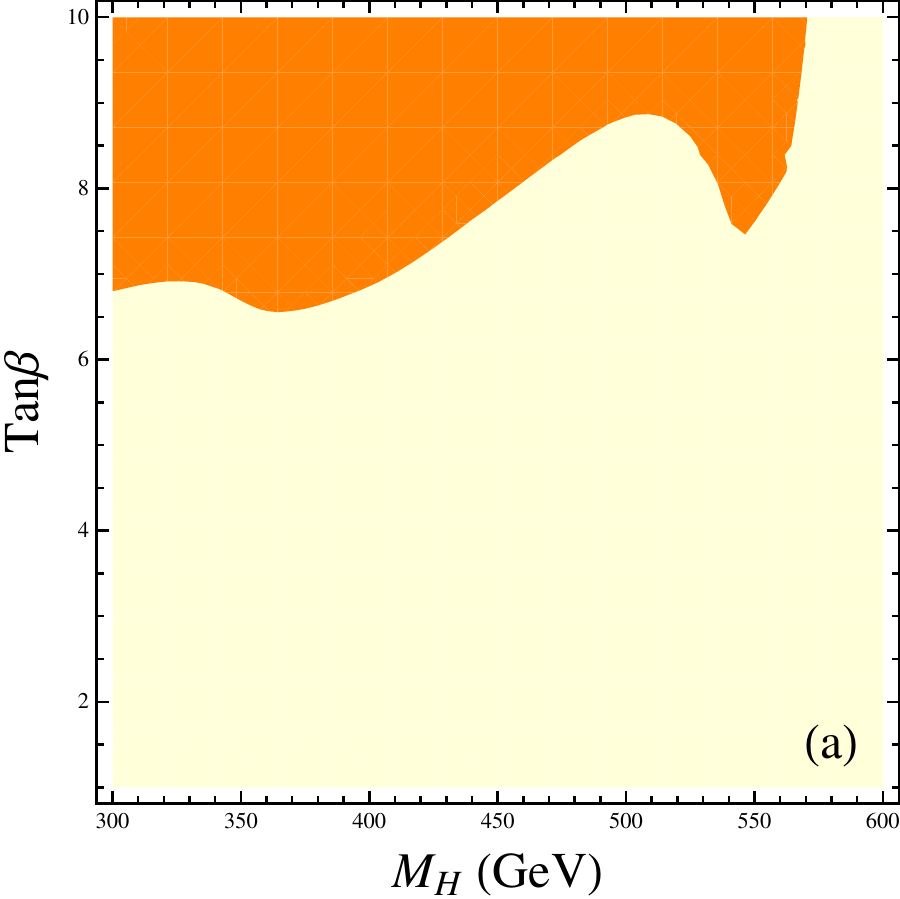}
\includegraphics[width=6.8cm,height=4.5cm]{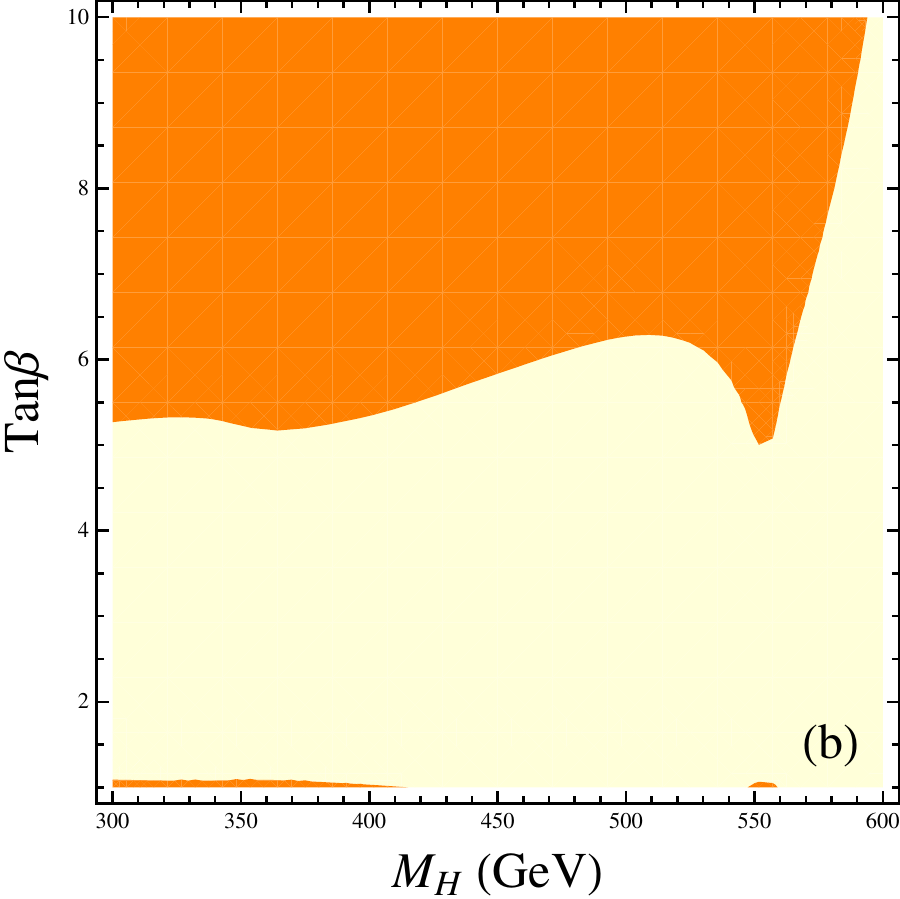}\\
\includegraphics[width=6.8cm,height=4.5cm]{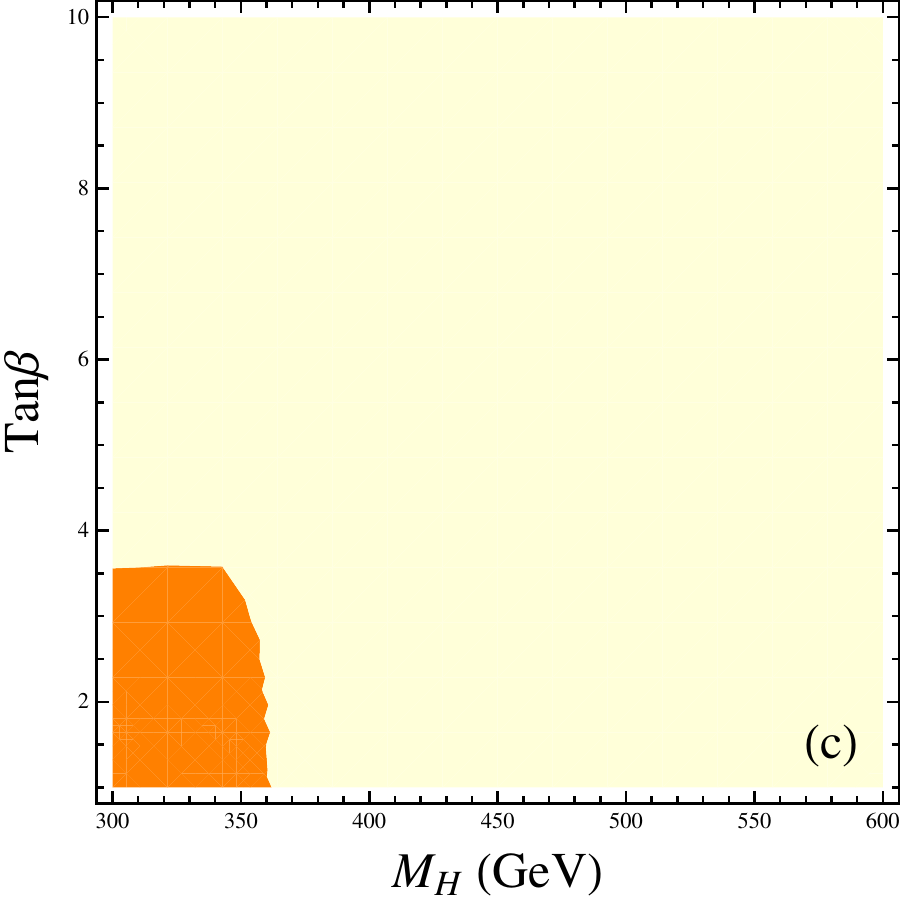}
\includegraphics[width=6.8cm,height=4.5cm]{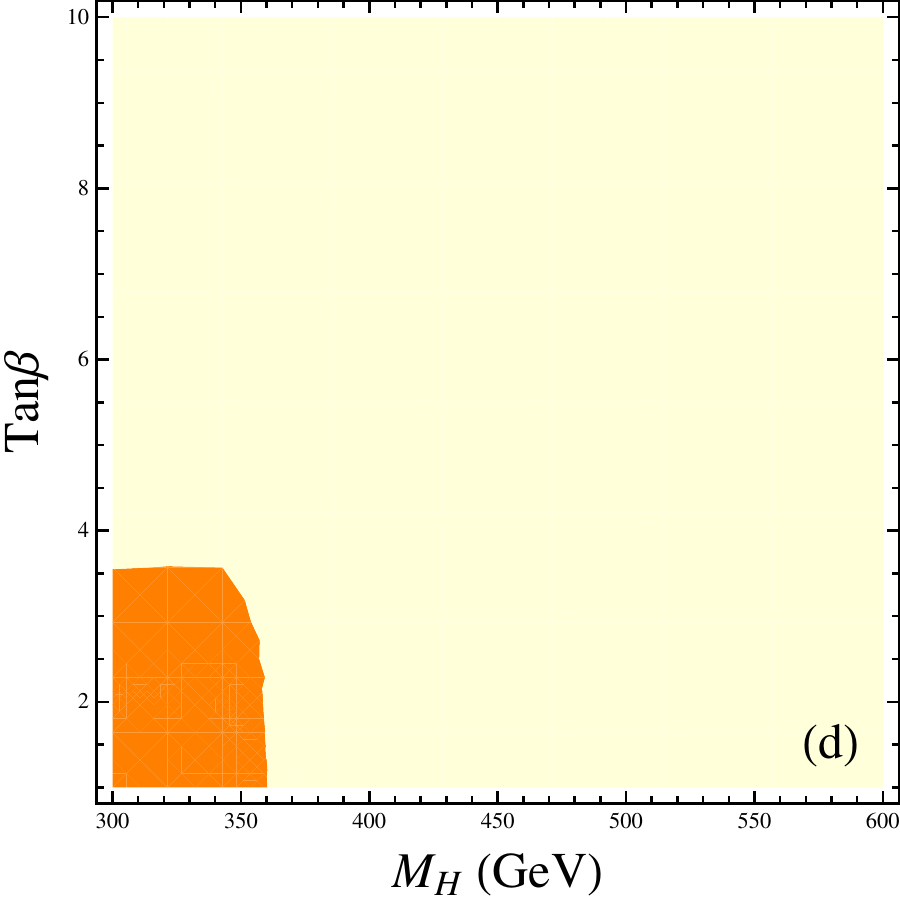}\\
\caption{\label{fig:MHtb_reach} The signal reaches for the $H\to b\bar b\gamma\gamma$ searches on the $(M_H\,,\tan\beta)$ plane. Upper left: 2HDM-I for $\int \mL dt=1000\,\fb^{-1}$. Upper right: 2HDM-I for $\int \mL dt=3000\,\fb^{-1}$. Lower left: 2HDM-II for $\int \mL dt=1000\,\fb^{-1}$. Lower right: 2HDM-II for $\int \mL dt=3000\,\fb^{-1}$. Parameter regions of $(M_H\,,\tan\beta)$ in orange color are within the reach for each case. The alignment parameters $c_{\beta-\alpha}$ are taken according to Eq.~(\ref{eq:align_para}). }
\end{figure}

Based on the cut-based analysis before, it is straightforward to further look at the LHC search potential to $H$ via the $b\bar b\gamma\gamma$ final states. For the LHC discovery, we require the number of signal events after the selection to satisfy $S\ge  {\rm max}\{ 5\sqrt{B}\,,10\}$. In Fig.~\ref{fig:H300_search}, we present the search sensitivities to the 2HDM $(c_{\beta-\alpha}\,,\beta)$ parameter space for the $M_{H}=300\,\GeV$ case. The search sensitivities are shown for the LHC run with $\int \mL dt=100\,\fb^{-1}$ (two upper panels) and $\int \mL dt=3000\,\fb^{-1}$ (two lower panels). In each plot, the yellow-shaded regions are within reach for the corresponding integrated luminosity case. Altogether, we impose the current global fit to the 2HDM parameters $(c_{\beta-\alpha}\,,\beta)$ from the light CP-even Higgs boson signal strengths by following Ref.~\cite{Craig:2013hca} with the green and brown bends representing the $68\,\%$ and $95\,\%$ C.L., respectively. For the LHC run 2 accumulating the integrated luminosity to $\sim 100\,\fb^{-1}$, one could already probe the $M_{H}=300\,\GeV$ case at the low-$t_{\beta}$ region for 2HDM-I and 2HDM-II. Significant improvements are shown at the large-$t_{\beta}$ region with the HL LHC runs up to $\int \mL dt=3000\,\fb^{-1}$. For the $M_{H}=300\,\GeV$ in both 2HDM-I and 2HDM-II, the whole parameter space allowed by the current global fit of $125\,\GeV$ Higgs is almost within reach via the $b\bar b\gamma\gamma$ channel when the integrated luminosity is up to $\sim 3000\,\fb^{-1}$.

We proceed to explore the search sensitivities in the large mass regions by restricting the parameter $c_{\beta-\alpha}$ along with the alignment limits according to Eq.~(\ref{eq:align_para}). Cross sections for the inclusive signal processes $\sigma[pp\to HX]\times {\rm Br}[H\to hh\to b\bar b \gamma\gamma ]$ are evaluated together with the kinematic cut efficiencies taken into account for different $M_{H}$ inputs. By imposing the same set of kinematic cuts to the SM background processes following each individual signal sample, we count the number of events left. In Fig.~\ref{fig:Hmass_reach}, we demonstrated the mass reach for $H$ via the $b\bar b\gamma\gamma$ final states with various $t_{\beta}$ inputs for 2HDM-I and 2HDM-II. It turns out for the 2HDM-I case with the large $t_{\beta}$ input, the LHC $14\,\TeV$ runs with $\int \mL dt$ up to $1000\,\fb^{-1}-3000\,\fb^{-1}$ can almost probe the $M_H\in (300\,,600)\,\GeV$ range via the $b\bar b\gamma\gamma$ final states. For the 2HDM-II case with the low-$t_{\beta}$ input, the mass reach is roughly below $2m_{t}$ with the full HL LHC runs, in accordance with the decay branching ratios shown in Fig.~\ref{fig:HBR_mass}.

 Further illustrations of the signal reaches are performed on the $(M_H\,,t_\beta)$ plane with the fixed alignment parameters of $c_{\beta-\alpha}$ for 2HDM-I and 2HDM-II, as displayed in Fig.~\ref{fig:MHtb_reach}. For the 2HDM-I case, increasing $\int \mL dt$ from $1000\,\fb^{-1}$ up to $3000\,\fb^{-1}$ would also reach the parameter regions with smaller inputs of $t_\beta$. Meanwhile, it is noticed that searches for the $H\to b\bar  b \gamma\gamma$ decay mode are of minor attraction for the 2HDM-II case with the large $M_H$ inputs in both small-$t_\beta$ and large-$t_\beta$ regions. This was expected in the signal evaluations for the $H\to b\bar b\gamma\gamma$ channel as shown in Fig.~\ref{fig:Hbbgaga_sig}. On the other hand, the conventional searches for the heavy CP-even Higgs were performed in the MSSM scenario at both LEP~\cite{Schael:2006cr} and LHC~\cite{CMS:2013hja}. The recent LHC searches~\cite{CMS:2013hja} were performed for the conventional $H\to \tau^+\tau^-$ mode, which are particularly relevant for the heavy CP-even Higgs searches at the large-$t_\beta$ regions. It is expected that the upcoming LHC runs at $14\,\TeV$ could further probe the heavy CP-even Higgs at the large-$t_\beta$ region through the $\tau^+\tau^-$ final states beyond the current mass limits.


\section{Conclusion and Discussion}
\label{section:conclusion}

In this work, we suggested that searches for the $b\bar b\gamma\gamma$ final states of a heavy CP-even Higgs in the 2HDM can be considered as a potentially promising channel for the upcoming LHC runs at $14\,\TeV$. Such states are due to the possible $H\to hh$ decay modes from the 2HDM Higgs potential. Within the framework of the general 2HDM, one is free to set the heavy Higgs boson masses and quartic couplings subject to the theoretical constraints. With proper parameter choices, it is possible to enhance the CP-even Higgs cubic coupling term $\lambda_{Hhh}$; hence, the $H\to hh$ decay mode becomes the most dominant one over a broad mass range of $250\,\GeV\lesssim M_{H}\lesssim 600\,\GeV$. To search for the final states with two $125\,\GeV$ Higgs bosons, we considered the combination of $b \bar b\gamma\gamma$ as our priority. Such a combination was also regarded as a priority for the SM Higgs boson self-coupling measurement, due to the manageable SM background contributions. By performing a cut-based analysis of different samples of $M_{H}$ inputs, we obtained the most optimal cuts and imposed them sequentially for both signal and background processes. For a hypothetical heavy CP-even Higgs boson with mass of $M_{H}=300\,\GeV$, the LHC runs at $14\,\TeV$ could probe much of the 2HDM parameter space allowed by the current global fit to $125\,\GeV$ Higgs signal strengths. We also discussed the mass reaches for $H$ via the $b\bar b\gamma\gamma$ final states, with the 2HDM parameters restricted along with the alignment limit. Depending on the 2HDM setups on their Yukawa sectors, the mass reach can be up to $\sim 600\,\GeV$ for the 2HDM-I with large-$t_{\beta}$ inputs at the HL LHC runs. On the other hand, the 2HDM-II case is generally challenging at the high mass region, given the suppression to the $H\to hh$ decay modes after the $t \bar t$ threshold.

More generally, the decay mode of $H\to hh$ exists in models with an extended Higgs sector in addition to the 2HDM case. Given that the signal strength of $125\,\GeV$ Higgs boson is close to the SM predictions with the current data, the future experimental searches for the extra heavy scalars decaying to a pair of SM-like Higgs bosons can be considered in a general sense. For the $b\bar b\gamma\gamma$ we focused on here, it is straightforward to suppress the SM background contributions by the kinematic cuts we analyzed above. Additionally, the search sensitivities via decay modes such as $b\bar b b \bar b$, $b \bar b W^{+}W^{-}$, and $b\bar b \tau^+ \tau^-$ can be further studied by the sophisticated jet substructure analysis. Combing the searches through various channels, it is likely to probe the extra heavy scalar in extended Higgs sectors at the upcoming LHC 14 TeV runs.


\section*{Acknowledgments}

We would like to thank Hongjian He for the initial motivation and very fruitful discussions for this work. We also would like to thank Jinmian Li, Ye Li, Jia Liu, Tao Liu, Yandong Liu, Olivier Mattelaer, and Qishu Yan for very useful discussions and communications during the preparation of this work. NC and LCL are partially supported by the NSF of China (under grants 11275101, 11135003) and National Basic Research Program (under grant 2010CB833000). LCL is also supported by the Graduate Student Overseas Exchanging Program of Tsinghua University. CD and YQF are partially supported by the Thousand Talents Program (under grant Y25155AOU1).

\appendix

\section{test}

\end{document}